\documentclass[12pt]{iopart}

\usepackage{iopams}
\usepackage{color}
\usepackage{caption}
\usepackage{verbatim}
\usepackage{graphicx}
\usepackage{ulem}
\usepackage{subfig}

\usepackage{color}

\def\@{\hskip.8pt}
\def\?{\hskip.3pt}
\def\plus#1#2{\vrule height#1pt width0pt depth#2pt}

\def\mbf#1{\mathbf{#1}}

\def \red#1 {\textcolor{red}{#1}}
\def \blue#1 {\textcolor{blue}{#1}}

\begin{document}
\title[Where Bayes tweaks Gauss]{Where Bayes tweaks Gauss: Conditionally Gaussian priors for stable multi-dipole estimation}
\author{Alessandro Viani$^1$, Gianvittorio Luria$^1$, Harald Bornfleth$^2$
and Alberto Sorrentino$^1$}

\address{$^1$ Department of Mathematics, University of Genoa, Via Dodecaneso $35$,\\  $16146$ Genoa, Italy}
\address{$^2$ BESA GmbH, Freihamer Str. 18, 82166 Gr\"{a}felfing, Germany}
\ead{sorrentino@dima.unige.it}
\begin{abstract}
We present a very simple yet powerful generalization of a previously described model and algorithm for estimation of multiple dipoles from magneto/electro-encephalographic data. 
Specifically, the generalization consists in the introduction of a log-uniform hyperprior on the standard deviation of a set of conditionally linear/Gaussian variables. 
We use numerical simulations and an experimental dataset to show that the approximation to the posterior distribution remains extremely stable under a wide range of values of the hyperparameter, virtually removing the dependence on the hyperparameter. 
\end{abstract}

\ams{ 62F15,92C55, 65C05}
\noindent{\it Keywords\/}: Bayesian inverse problems, hyperprior, M/EEG, conditionally Gaussian models, sequential Monte Carlo

\submitto{\IP}

\maketitle
\section{Introduction}

MagnetoEncephaloGraphy (MEG) and ElectroEncephaloGraphy (EEG) are non--invasive functional neuroimaging techniques that provide measurements of 
the electromagnetic brain activity at selected locations on the scalp \cite{hamalainen1993magnetoencephalography,niedermeyer2005electroencephalography}. The very high sampling rate of their recordings, which can exceed 1 kHz, 
is the key to the study of the dynamics of neural processes with very high precision in time. The localization of the active regions in the brain from M/EEG data is accomplished through the solution of the neuroelectromagnetic inverse problem which consists in estimating the neuronal primary current distribution generating the measured electromagnetic field \cite{sa87}. Unfortunately, this problem is known to be ill-posed, suffering from non-uniqueness of the solution in the absence of constraints. As a consequence, providing the correct prior knowledge becomes crucial.

For the inverse problem to be tackled, it is required that the forward problem is properly set and solved first: this amounts to giving a parametric representation of the sources, and to exploit the information about the  physical and geometrical properties of the head (which can be gathered from high resolution anatomical Magnetic Resonance Imaging) for modeling how the electromagnetic field propagates through the brain compartments. Two main source models have been proposed in the literature. In the former, the activity of the neuronal population inside a little brain volume is represented by a point source, called equivalent current dipole (ECD), and the whole brain activity is explained by a small --- but a priori unknown --- number of ECDs. Each ECD will typically have a location that remains fixed across relatively large temporal intervals, and a moment which may vary depending on how the neural populations within that specific domain activate and synchronize. 
The latter model, called the Distributed Source (DS) model, assumes instead the neural current to be a continuous vector field inside the brain volume, which is discretized for computational purposes.

From a mathematical perspective, the forward ECDs model is nonlinear in dipole locations, which makes it harder to invert than the DS model which is basically linear.
However, the ECDs model has at least two advantages with respect to the DS model. On the one hand, it is easily interpretable\@: summarizing the whole brain
activity with few ECDs provides a clear-cut answer to the localization problem, which can be particularly useful in clinical scenarios such as the pre-surgical evaluation of epileptic patients \cite{knake2006value}.
 On the other hand, being it an inherently parsimonious model involving only a small set of parameters, 
the ECDs model allows for a fully Bayesian approach in which an approximation of the full posterior probability distribution is attained and provides uncertainty quantification of the solution;
this again can turn out to be useful in clinical scenarios where decision making can also be based on estimated uncertainty.

In a couple of recent publications \cite{sorrentino2014bayesian, sommariva2014sequential} a Sequential Monte Carlo (SMC) algorithm has been proposed to solve the ECDs estimation problem in a fully Bayesian setting. In the first work, an adaptive SMC algorithm was proposed to sample the whole posterior distribution for the ECD configuration, given a single spatial distribution of the M/EEG data (also called \textit{topography}). The second work extended the first one by allowing estimation of ECD configurations from a whole sequence of M/EEG topographies, under the assumption that the dipole locations remained fixed. 
In order to make the computation feasible, a Gaussian prior for the source strength was assumed: this allowed to exploit the linear/Gaussian substructure of the problem, so that Monte Carlo sampling was used for the variables on which data depend non-linearly, namely the number of ECDs and their locations, while the conditionally Gaussian posterior for the dipole moments was computed analytically.

However, in most circumstances, 
the Gaussian prior happens to be relatively strong,
typically not reflecting any actual prior knowledge and possibly jeopardizing the correct identification of brain activity.
Indeed, as shown in \cite{sommariva2014sequential}, too small a value of the prior width tend to result in overestimation of the number of sources, because multiple weak dipoles are needed to explain the data produced by a single, strong source; and too big a value may result in underestimation of the number of sources, or in a bias in the source location towards deeper sources.

In the present work we extend the model \cite{sommariva2014sequential} by introducing a hyperprior on the prior width $\sigma_q$, 
thus better representing the little prior information we have on source strength. 
This new model has the advantage of providing a marginal prior distribution which is no longer Gaussian and has a much wider support, while leaving intact the conditionally 
linear/Gaussian structure that allows processing of multiple topographies with a feasible computational cost. 
As we will show by numerical simulations, the proposed model becomes very robust against mis-specifications of the value of the hyperparameter, even by two orders of magnitude. 
From the computational point of view, $\sigma_q$ is considered as a free parameter and is sampled with the same SMC technique as the other ones, involving a negligible additional computational cost. 



The paper is organized as follows: In Section 2 we summarize the model and algorithm described in \cite{sorrentino2014bayesian} and \cite{sommariva2014sequential}. In Section 3 we provide the generalization of the model through the use of a hyperprior. In Section 4 the hierarchical model is tested and compared against the previous one using synthetic MEG and EEG data. In Section 5 the same comparison is performed on a set of experimental MEG data publicly available. Finally, in Section 6 we offer our conclusions.

\section{Bayesian multi--dipole localization in M/EEG with SESAME}\label{no_hyper_section}

In this Section we briefly review the mathematical model and algorithm proposed in \cite{sorrentino2014bayesian,sommariva2014sequential}, which the current work is based on. For more details we invite the reader to refer to the original articles. The algorithm was subsequently included within the BESA Research software\footnote{\texttt{https://www.besa.de/products/besa-research/new-features-in-besa-research-7-0/}} and we shall henceforth call it by its commercial name SESAME, which is an acronym for SEquential Semi Analytic Montecarlo Estimation; SESAME is also available as free and open source code\footnote{\texttt{https://github.com/pybees/sesameeg}}. 

\subsection{Mathematical model}
An ECD is a mathematical representation of an electrical current concentrated at a single point in space.
Given a brain discretization, we call $r$ the index representing the location of the ECD in the discretized grid and 
$\mbf{q}$ the 3-dimensional vector representing the direction and the intensity of the current in a given frame of reference.
Denoting by $n_s$ the number of sensors in the M/EEG device, a single ECD produces a spatial distribution of M/EEG data $\mbf{y} := (y_1, \ldots, y_{n_s})$ that 
can be represented compactly as 
\begin{equation}\label{eq:exact_field_1_dip}
 \mbf{y} = G(r) \cdot \mbf{q}\@, 
\end{equation}
where $G(r)$ is the forward operator computed at the ECD location $r$. 
Equation (\ref{eq:exact_field_1_dip}) shows that the data depend linearly on $\mbf{q}$ and non-linearly on $r$ through  $G$.

The forward operator, often referred to as the \textit{leadfield matrix} in the M/EEG literature, defines the relationship between source parameters and observed M/EEG data; the leadfield matrix is computed on the basis of the M/EEG device sensors configuration and of a propagation model for the electromagnetic field which relies on a given geometric and physical model of the head, including the shapes of various compartments (brain, cerebrospinal fluid, skull, scalp) and their respective magnetic/electrical conductivities. 
In this work, $G$ is assumed to be known and we do not provide further details, which can be found in many references \cite{mosher1999eeg,acar2010neuroelectromagnetic,pursiainen2011forward}.

In multi-dipole models, M/EEG data are given by the superposition of the electromagnetic fields generated by the single dipoles. 
Let $\mbf{y}_{1:T}$\vspace{1pt} be a $T$-sample-long window of a M/EEG recording 
during which both the number of ECDs $n_D$ as well as their locations $r^{\@1:n_D}$ are supposed to be 
fixed\@\footnote{Henceforth, given any variable $x$, we shall make use of the following compact notation: 
\[
x_{\@A:B} := \{x_{\@i}\}_{\/i=A, \ldots, B}\quad ,\quad x^{\@A:B} := \{x^{\@i}\}_{\/i=A, \ldots, B}\ .
\]
}.
Assuming each topography $\mbf{y}_t$ to be affected by a corresponding zero--mean Gaussian additive noise $\mbf{n}_t$ and denoting by 
$\mbf{q}^{\@d}_{\@1:T}$  the course of the $d$--th ECD 
moment, the forward model can be expressed as\@:
\begin{equation}\label{eq:obs_time}
\mbf{y}_t = \sum_{d=1}^{n_D} G(r^{d}\@) \cdot \mbf{q}^{\@d}_{\@t} +\ \mbf{n}_t\@, \qquad \qquad t = 1, \ldots, T\@.
\end{equation}

We observe incidentally that the model in (\ref{eq:obs_time}) represents an ECD model only as long as $n_D$ takes small values and the ECD locations are not known a priori. If, on the contrary, we let $n_D$ take large values (in the order of $10^4$), and the index $r^d$ run through all possible locations in the brain discretization, then eq. (\ref{eq:obs_time}) represents a distributed model. In this sense, the ECD model can be seen as an extremely sparse version of the distributed model, and the present work relates strongly to, e.g., \cite{gretal12}.

\subsection{Statistical model}

In a Bayesian setting  \cite{evans2002inverse,kaipio2006statistical}, all the involved variables are interpreted as realizations of corresponding random 
variables, and it is assumed that their probability distributions can be expressed in terms of probability densities. In the following we shall denote random variables by capital letters and their realizations by lowercase letters.
In the present work $\mbf{Y}_t$, $\mbf{Q}^{\@1:n_D}_{\@t}$, and $\mbf{N}_t$ are assumed to be continuous variables while $N_D$ and $R^{\@1:n_D}$ are considered discrete variables.

Given a collection of measured data $\mbf{y}_{1:T}$, the posterior distribution as given by Bayes theorem is

\begin{equation}
    \pi(\mbf{q}_{1:T}^{1:n_D},r^{1:n_D},n_D|\mbf{y}_{1:T})= \frac{\pi(\mbf{y}_{1:T}|r^{1:n_D},\mbf{q}_{1:T}^{1:n_D},n_D) \pi(r^{1:n_D},\mbf{q}_{1:T}^{1:n_D},n_D)}{\pi(\mbf{y}_{1:T})}~~~,
    \label{eq:Bayes1}
\end{equation}
where: $\pi(\mbf{y}_{1:T}|r^{1:n_D},\mbf{q}_{1:T}^{1:n_D},n_D)$ is the likelihood of a given set of parameters, and is assumed to be a Gaussian distribution; $\pi(r^{1:n_D},\mbf{q}_{1:T}^{1:n_D},n_D)$ is the prior of the same set of parameters, which are assumed to be independent on each other; the denominator is a normalization constant, usually unknown.
Because the number $n_D$ of ECD is unknown, the support of this distribution is a variable dimension space, with potentially high-dimensional subsets depending on how large $n_D$ and $T$ are. As a consequence, inference from this posterior distribution is not straightforward, and efficient computational strategies are needed to explore it. In \cite{sorrentino2014bayesian} an adaptive SMC algorithm \cite{del2006sequential} was proposed to approximate the full posterior in the computationally less demanding case $T=1$ . 

However, we notice that the posterior distribution naturally splits as follows

\begin{equation}
\pi(\mbf{q}_{1:T}^{1:n_D},r^{1:n_D},n_D|y_{1:T})= \pi(\mbf{q}_{1:T}^{1:n_D}|\mbf{y}_{1:T},r^{1:n_D},n_D) \; \pi(r^{1:n_D},n_D|\mbf{y}_{1:T})~~~.
\label{eq:split1}
\end{equation}
In \cite{sommariva2014sequential} this splitting was used to devise an efficient approximation algorithm to treat the case $T>>1$: under the additional assumptions that dipole moments $\mbf{Q}_{1}^{1},\dots,\mbf{Q}_{T}^{n_D}$ are independent on each other and with a Gaussian distribution, the first term at the right hand side can be computed analytically and only $\pi(r^{1:n_D},n_D|\mbf{y}_{1:T})$ requires Monte Carlo sampling.
More formally:
\begin{itemize}
    \item the prior distribution of the dipole moments is assumed to be
\begin{equation}
\pi(\mbf{q}_{1:T}^{1:n_D}) = \prod_{t=1}^T\prod_{d=1}^{n_D}\mathcal{N}\bigl(\mbf{q}_t^d;0,\Gamma_{q}\bigl)
\label{old_prior}
\end{equation}
where $\mathcal{N}(\cdot;\mu,\Gamma)$ denotes the Gaussian distribution of mean $\mu$ and covariance matrix $\Gamma$, and $\Gamma_{q}= diag({{\sigma}}_{q}^2, {\sigma}_q^2, {{\sigma}}^2_{q})$;
    \item the likelihood is assumed to be
\begin{equation}
\eqalign{
    \pi(\mbf{y}_{1:T}|\mbf{q}_{1:T}^{1:n_D},r^{1:n_D},n_D) & = \prod_{t=1}^T \pi(\mbf{y}_t|\mbf{q}_t^{1:n_D},r^{1:n_D},n_D) = \\ & =
    \prod_{t=1}^T \mathcal{N}(\mbf{y}_t; \sum_{d=1}^{n_D}G(r^d)\cdot \mbf{q}_t^d; \Gamma_N)~~~,}
\end{equation}
where $\Gamma_N$ is the noise covariance matrix, which is assumed to be known.
\end{itemize}

Under these two assumptions, it is possible to prove that 
\begin{itemize}
\item  the marginal likelihood $\pi\/(\mbf{y}_{1:T}\,|\,n_D, r^{\@1:n_D})$ is:
\begin{equation}\label{eq:likelihood_marginal}
 \pi\/(\mbf{y}_{1:T}\,|\,n_D, r^{\@1:n_D})\ =\  \prod_{t=1}^T\ \mathcal{N}(\mbf{y}_t; \mbf{0},  \Gamma_l)\ , 
\end{equation}
with\@\footnote{Given any matrix $M$, we denote by ${}^t\!\@M$ its transpose.}  $\Gamma_l = {\sigma}_q^2\  G\/(r^{\@1:n_D})\  {}^t\!\@G\/\left(r^{\@1:n_D}\right) +\, \Gamma_N\,$.
\item the conditional posterior $\pi\/(\mbf{q}^{\@1:n_D}_{\@1:T}\,|\,\mbf{y}_{1:T}, n_D, r^{\@1:n_D})$ is

\begin{equation}\label{eq:post_q_density}
\eqalign{
\pi\/(\mbf{q}^{\@1:n_D}_{\@1:T}\,|\,\mbf{y}_{1:T}, n_D, r^{\@1:n_D}) &= \prod_{t=1}^T\ \pi\/(\mbf{q}^{\@1:n_D}_{\@t}\,|\,\mbf{y}_{t}, n_D, r^{\@1:n_D})\\ 
 &= \prod_{t=1}^T\ \mathcal{N}\bigl(\mbf{q}^{\@1:n_D}_{\@t};\bar{\mbf{q}}^{\@1:n_D}_{\@t}, \Gamma_t)
}
\end{equation} 
where
\numparts
\begin{equation}\label{eq:post_q}
\bar{\mbf{q}}^{\@1:n_D}_{\@t} = {\sigma}^2_q\@ \ {}^t\!\@G\/(r^{\@1:n_D})\, {\Gamma_l}^{-1}\@ \mbf{y}_{t}
\end{equation}
and
\begin{equation}
\Gamma_t = {\sigma}_q^2\,\mathbf{I}_{\plus60 3\@n_D} -\, {\sigma}^4_q\ {}^t\!\@G\/(r^{\@1:n_D})\, {\Gamma_l}^{-1}\@G\/(r^{\@1:n_D})\,.
\end{equation}
\endnumparts
\end{itemize}

As for the latter factor in the right--hand side of equation (\ref{eq:split1}), it can be expressed by Bayes' theorem as
\begin{equation}    \label{eq:Bayes3}
\pi\/(n_D, r^{\@1:n_D}\@|\,\mbf{y}_{1:T})\ =\ \frac{\pi\/(\mbf{y}_{1:T}\,|\,n_D, r^{\@1:n_D})\ \pi\/(n_D, r^{\@1:n_D})}{\pi(\mbf{y}_{1:T})}~~~,
\end{equation}
where $\@\pi\/(\mbf{y}_{1:T}\,|\,n_D, r^{\@1:n_D})$ is the marginal likelihood (\ref{eq:likelihood_marginal}),  $\pi\/(n_D, r^{\@1:n_D})$ is the marginal prior, and 
the denominator is a normalizing constant.

\subsection{The SESAME algorithm}

The SESAME algorithm  was devised in \cite{sommariva2014sequential} for performing source modeling from multiple M/EEG topographies.
SESAME exploits the semi--linear structure of the model (\ref{eq:obs_time}) and approximates the full posterior probability distribution (\ref{eq:split1}) in two steps: 
first, it makes use of an Adaptive Sequential Monte Carlo (\/ASMC\/) sampler to approximate the
distribution $\pi\/(n_D, r^{\@1:n_D}\@|\,\mbf{y}_{1:T})$ of the posterior probability of those parameters on which data depend nonlinearly; 
then, the mean and covariance matrix of the conditional distribution $\pi\/(\mbf{q}^{\@1:n_D}_{\@1:T}\,|\,\mbf{y}_{1:T}, n_D, r^{\@1:n_D})$ are analytically computed 
through formulas  (\ref{eq:post_q},\@\textit{b}).\vspace{1pt}

We now provide a brief summary of the ASMC sampler. Sequential Monte Carlo samplers were originally described in \cite{del2006sequential}: their construction is relatively complex and will not be repeated here; below we just give the main ideas and the computational recipe, in order to make the results reproducible.

The algorithm aims at obtaining a weighted sample that can be used to approximate integrals of the posterior distribution. Three main ideas underlie the ASMC algorithm. 

First, the posterior distribution is not sampled directly: instead, a sequence of auxiliary distributions 
\begin{equation}\label{eq:sequence}
\left\{ \pi_i\/(n_D, r^{\@1:n_D}\@|\,\mbf{y}_{1:T}) \propto \pi\/(\mbf{y}_{1:T}\,|\,n_D, r^{\@1:n_D})^{\@\alpha(i)}\, \pi\/(n_D, r^{\@1:n_D}) \right\}_{i=1,\dots,n_I}
\end{equation}
is defined, with $\alpha(1)=0$, $\alpha(n_I)=1$ and $\alpha(1) < \alpha(2) < \ldots < \alpha(n_I)$.
Notice that the sequence starts with the prior distribution and ends with the target posterior distribution.

\smallskip
Second, importance sampling and Markov Chain Monte Carlo techniques are combined to approximate sequentially each element of (\ref{eq:sequence}) 
with the weighted sample set 
$\left\{\@x_i^{\@p}  := \big( n_D, r^{\@1:n_D}\big)_i^{\@p}\ ;\, w_i^{\@p}\@\right\}_{p=1,...,n_P} \@;$
each sample $x_i^{\@p}$, or \textit{particle}, is a point in the hypothesis space whose coordinates represent a realization of the unknown random variables, 
namely of the number of active sources and of their location. The number of particles $n_P$ represents, roughly speaking, the number of candidate solutions that are 
tested in the Monte Carlo procedure.

\smallskip
Finally, the sequence of exponents $\{\alpha(i)\}_{\/i = 1, \ldots,  n_I}$, which determines the elements of (\ref{eq:sequence}), is not established a priori, 
but adaptively determined at run--time.
This means that the actual number of iterations $n_I$ is also determined online, even if it is always kept within given lower and upper bounds.\\

The algorithm works as follows.
At $i = 1$, the exponent $\alpha\/(1)$ is set to $0\@$; 
the initial sample set $x_1^{1:n_P} $  is drawn from the prior distribution and to each sample is assigned an uniform weight
$w_1^{\@p} = \frac{1}{n_P}\@\vspace{1pt}$. Subsequently, the following steps are iterated until $\alpha\/(i)$ reaches $1$:

\begin{itemize}
\item the next sample set   $x_i^{1:n_P} $ is obtained by drawing each particle from a $\pi_i$-invariant MCMC kernel centered in the current state 
$K_i(\ \cdot\, , x_{i-1}^{\@p})$; such kernel is obtained as the product of a Reversible Jump Metropolis--Hastings kernel \cite{green1995reversible}, 
accounting for a possible change in the number of dipoles in the particle, and $(n_D)_i^{\@p}$ Metropolis--Hastings kernels \cite{hastings1970monte}, 
for dipole locations evolution. This way, each particle explores the state space by allowing both the number of dipoles as well as their locations to change.

The increment or decrement by one of $(n_D)_i^{\@p}$ is attempted with probability of $\frac{1}{3}$ and of $\frac{1}{20}$, respectively. 
If a birth move is accepted, the location of the newborn dipole is uniformly distributed. 
If a death move is accepted, the excluded dipole is uniformly chosen among the existing ones.

As far as source locations are concerned, each dipole is let move only to a restricted neighbouring set of brain points, 
with a probability decreasing with the distance. 

\item $\alpha(i+1)$ is determined adaptively by bisection in such a way that the distance between $\pi_{i+1}$ and $\pi_i$ is within a prescribed interval; 
the distance between sampled distributions is measured by means of the ratio ${ESS\/(n+1)}/{ ESS\/(n)}$ where the Effective Sample Size (\/ESS\/) is defined as 
\begin{equation*}
 ESS\/(i) = \left[\ \sum_{p=1}^{n_P}\@ \left(w_i^{\@p}\right)^2\ \right]^{-1}
\end{equation*}
and the weights are given by 
\begin{equation*}
w_{i+1}^{\@p}\ = \frac{\tilde w_{i+1}^{\@p}}{\sum_{j=1}^{n_P}\tilde w_{i+1}^{\@j} }\@ ,
\end{equation*}
with
\begin{equation*}
\tilde w_{i+1}^{\@p}\ =\ w_{i}^{\@p}\, \frac{\pi_{i+1}\big(x^{\@p}_i\,|\,\mbf{y}_{1:T}\big)}{\pi_{i}\big(x^{\@p}_i\,|\,\mbf{y}_{1:T}\big)}\,.
\end{equation*}

The exponent $\alpha(n+1)$ is chosen in such a way that the ratio $\frac{ESS\/(i+1)}{ ESS\/(i)}$ falls between $0.9$ and $0.99$.

\item whenever the ESS falls below $\frac{n_P}{2}$, a systematic resampling step \cite{douc2005comparison} is applied in order to prevent all but one sample from having negligible weights.
\end{itemize}

\section{A hyperprior for SESAME}
\label{hyper_section}

We now propose a simple yet powerful generalization of the model described in the previous section. Indeed, one of the main limitations of the previously described model is that the Gaussian prior for the dipole moments is a relatively strong prior, that was introduced to make the algorithm computationally feasible but typically misrepresents the available prior information on the source strength. 
From a practical perspective, the main drawback is that the posterior distribution might depend strongly on the value of the hyperparameter $\sigma_q$: in \cite{luria2019bayesian} the authors showed that different values of $\sigma_q$ (also combined with different values of the hyperparameter of the prior of the number of sources) lead to source reconstructions characterized by different degrees of focality. In real scenarios it is not obvious what a good value of the hyperparameter might be; one possible solution is to  run the algorithm multiple times, with different values of the hyperparameter, and then compare the corresponding results either visually or in terms of their marginal likelihood. However, this approach is computationally demanding and does not overcome the limitation of a purely Gaussian prior that might lead, in the worst case, to a biased inference on the source locations.

Here we propose a simple solution, consisting in generalizing the statistical model by including the hyperparameter $\sigma_q$ among the unknowns, and assigning it a hyperprior.  Therefore we now make inference using the posterior distribution $\pi(\mbf{q}_{1:T}^{1:n_D},r^{1:n_D},n_D,\sigma_q|\mbf{y}_{1:T})$, that can be obtained by Bayes theorem as
\begin{equation}
	\hspace{-1.5cm}
    \pi(\mbf{q}_{1:T}^{1:n_D},r^{1:n_D},n_D,\sigma_q|\mbf{y}_{1:T})  = 
    \frac{\pi(\mbf{y}_{1:T}|\mbf{q}_{1:T}^{1:n_D},r^{1:n_D},n_D,\sigma_q) \pi(\mbf{q}_{1:T}^{1:n_D},r^{1:n_D},n_D,\sigma_q)}{p(\mbf{y}_{1:T})}
    \label{eq:Bayes2}
\end{equation}
We notice that the likelihood $\pi(\mbf{y}_{1:T}|\mbf{q}_{1:T}^{1:n_D},r^{1:n_D},n_D,\sigma_q)$ in eq. (\ref{eq:Bayes2}) is exactly the same that appeared in eq. (\ref{eq:Bayes1}), where the conditioning on $\sigma_q$ was implicit because its value was fixed. The main difference is that the prior distribution now includes also the hyperparameter $\sigma_q$. Following \cite{sommariva2014sequential},
we assume that dipole locations, dipole moments, and number of dipoles are a priori independent on each other so that the prior splits
\[
\eqalign{
\pi(\mbf{q}_{1:T}^{1:n_D},r^{1:n_D},n_D,\sigma_q) & = \pi(\mbf{q}_{1:T}^{1:n_D},r^{1:n_D},n_D|\sigma_q) \pi(\sigma_q) = \\
& = \pi(\mbf{q}_{1:T}^{1:n_D}|\sigma_q) \pi(r^{1:n_D}) \pi(n_D) \pi(\sigma_q)~~~.}
\]
and as far as the individual components are concerned: 
\begin{itemize}
    \item we assume to have no prior knowledge on the dipole locations, therefore $\pi(r^{1:n_D})$ is uniform in the brain, with the constraint that two dipoles cannot be in the same grid point;
    \item with the aim of favouring parsimonious models, we set a Poisson prior with small mean ($<1$) for the number of dipoles; 
    \item we set a conditionally Gaussian prior for the dipole moments, i.e. $\pi(\mbf{q}_{1:T}^{1:n_D}|\sigma_q)$ is Gaussian as before. 
\end{itemize}

When it comes to the hyperprior, in this study we will be using a log-uniform distribution to encode ignorance on the order of magnitude of the hyperparameter: a log-uniform prior is a uniform prior on the logarithm of the value, and therefore assigns the same probability to intervals in different orders of magnitude (for instance $P([1,10]) = P([10,100])$); its density is simply given by 
\begin{equation}
\pi(\sigma_q) = \left\{ 
\begin{array}{lc}
\displaystyle \frac{1}{\sigma_q} \hspace{2cm} \sigma_{min} < \sigma_q < \sigma_{max} \\
0 \hspace{2.3cm} \mbox{elsewhere} 
\end{array} 
\right.
\end{equation}
In order to define this hyperprior, we need to set the values of the two hyper-hyperparameters $\sigma_{min}$ and $\sigma_{max}$  (from here on referred to as hyperparameters, for simplicity). In the analysis below, we will be always using $\sigma_{max} = 10^3 \; \sigma_{min} $; covering three orders of magnitude is enough to account for within- and between-subject differences; this way, the hyperprior is defined by just one hyperparameter.

Two important observations are in order.
First, the new model can be considered as a generalization of the model in \cite{sommariva2014sequential}, that can be obtained from this one by using a hyperprior assigning all the probability to a single value $\sigma_q$.
Second, thanks to the introduction of the hyperprior, the marginal prior on the dipole moments $\pi(\mbf{q}_{1:T}^{1:n_D})$ is, in general, no longer Gaussian; specifically, our choice of a log-uniform hyperprior produces a marginal prior with fat tails, that assigns non-negligible prior probability to both large and small absolute values of the dipole moment. In practice, we overcome the classical limitation of Gaussian priors, i.e. the fact that they assign most of the probability mass to one single order of magnitude.


We now proceed to discuss how to approximate the posterior distribution in (\ref{eq:Bayes2}) through an adaptation of SESAME we will refer to as hierarchical-SESAME (h-SESAME) in the remainder of this article. First we observe that, again, the posterior splits as follows
\[
\pi(\mbf{q}_{1:T}^{1:n_D},r^{1:n_D},n_D,\sigma_q|\mbf{y}_{1:T})= \pi(\mbf{q}_{1:T}^{1:n_D}|r^{1:n_D},n_D,\sigma_q,\mbf{y}_{1:T}) \; \pi(r^{1:n_D},n_D,\sigma_q|\mbf{y}_{1:T})~~~,
\]
where $\pi(\mbf{q}_{1:T}^{1:n_D}|r^{1:n_D},n_D,\sigma_q,\mbf{y}_{1:T})$ is a Gaussian distribution of known mean and covariance matrix; in fact, it is the very same distribution that appeared in (\ref{eq:split1}), where the conditioning on $\sigma_q$ was implicit. Therefore, like in \cite{sommariva2014sequential}, sampling of the posterior is only needed for those variables on which data depend nonlinearly: the number of ECDs, their locations and the prior standard deviation. In practice, the only differences with respect to the algorithm sketched in the previous section are that
\begin{itemize}
    \item[(i)] each particle is now also determined by a value of $\sigma_q$, i.e. $x^i_n = (r^{1:n_D},n_D,\sigma_q)_n^i, w_n^i$
    \item[(ii)] at each iteration, an additional Metropolis-Hastings move is attempted on the value of $\sigma_q$: here we used a Gamma distribution as proposal distribution with shape parameter $k = 3$ and scale parameter $\theta=\frac{(\sigma_q)^p_i}{3}$, so that the proposal distribution has a mean $k\/\theta = (\sigma_q)^p_i$ and a relatively high variance
$k\/\theta^2$; we observed no major differences when using an inverse Gamma and a uniform distribution.
 \end{itemize}

Despite the simplicity of the proposed generalization, the numerical results in the next section provide strong evidence in favor of it.

\section{Numerical simulations}
\label{sec:num_sim}

We proceed to a numerical validation of h-SESAME and a comparison with plain SESAME.

\subsection{Synthetic data generation and processing}

We construct a total of 800 simulated datasets, as follows. 

\smallskip
We use the publicly available \texttt{sample} dataset\@\footnote{The dataset contains simultaneous M/EEG recordings together with subject's anatomical MRI scans. All data 
were acquired at MGH/HMS/MIT Athinoula A. Martinos Center Biomedical Imaging with a Neuromag Vectorview MEG device, a $60$-channel electrode  EEG cap and 
a Siemens $1.5$ T Sonata MRI scanner, using an MPRAGE sequence.}, taken from the MNE-Python package \cite{gramfort2013meg,gramfort2014mne}, to set a brain discretization and 
to compute two different forward solutions, one for MEG and the other for EEG, using a boundary element method and a surface segmentation computed with the watershed algorithm. 
The brain grid amounts to $18840$ points, the simulated MEG device has $305$ sensors ($203$ gradiometers and $102$ magnetometers) and the simulated EEG cap has $60$ electrodes.

For each of $n_D\@\in\@\{1,2,3,4\}$ and for each device we simulate $100$ different source configurations, by randomly drawing $n_D$ ECD locations and orientations, 
and assigning to each dipole the same time course, consisting of a $40$-sample-long bell-shaped curve.
To avoid making the ECDs too difficult to be discerned, we require that no two dipoles could be closer than $3\,$cm. The maximum value of each
simulated neuronal current is $200\,$nA\@m.

We separately compute $400$ MEG and $400$ EEG noise-free data, by projecting the source space currents to the sensor space using the corresponding forward operator.

We then proceed to adding noise. As far as the simulated MEG data is concerned, we add to each dataset a randomly-selected
interval of an empty room recording, which also has been taken from the MNE-Python \texttt{sample} dataset, and we then check that 
the signal--to--noise (SNR) ratio of the signal generated by each dipole is not lower than $3\,$dB. If not, the dipole si replaced by 
a newly simulated one. This process iterates until all ECDs fulfil the required SNR condition. As for the EEG data, white Gaussian noise is added with fixed standard deviation.
We notice that the effective SNRs of different datasets can be considerably different, as sources with different depths produce a different signal intensity.
\begin{figure}
\centering
\subfloat{\includegraphics[scale=0.40]{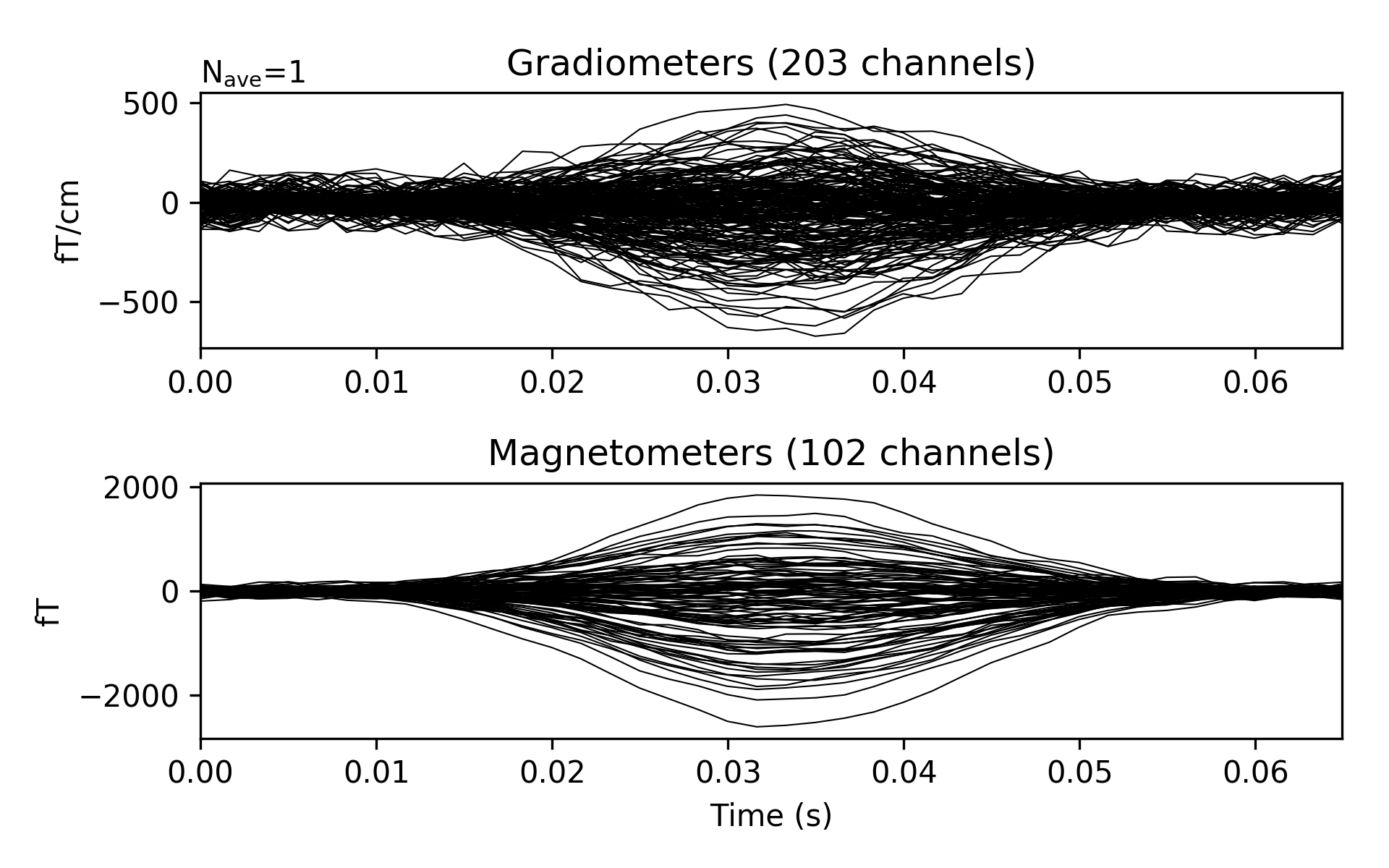}}
\subfloat{\includegraphics[scale=0.50]{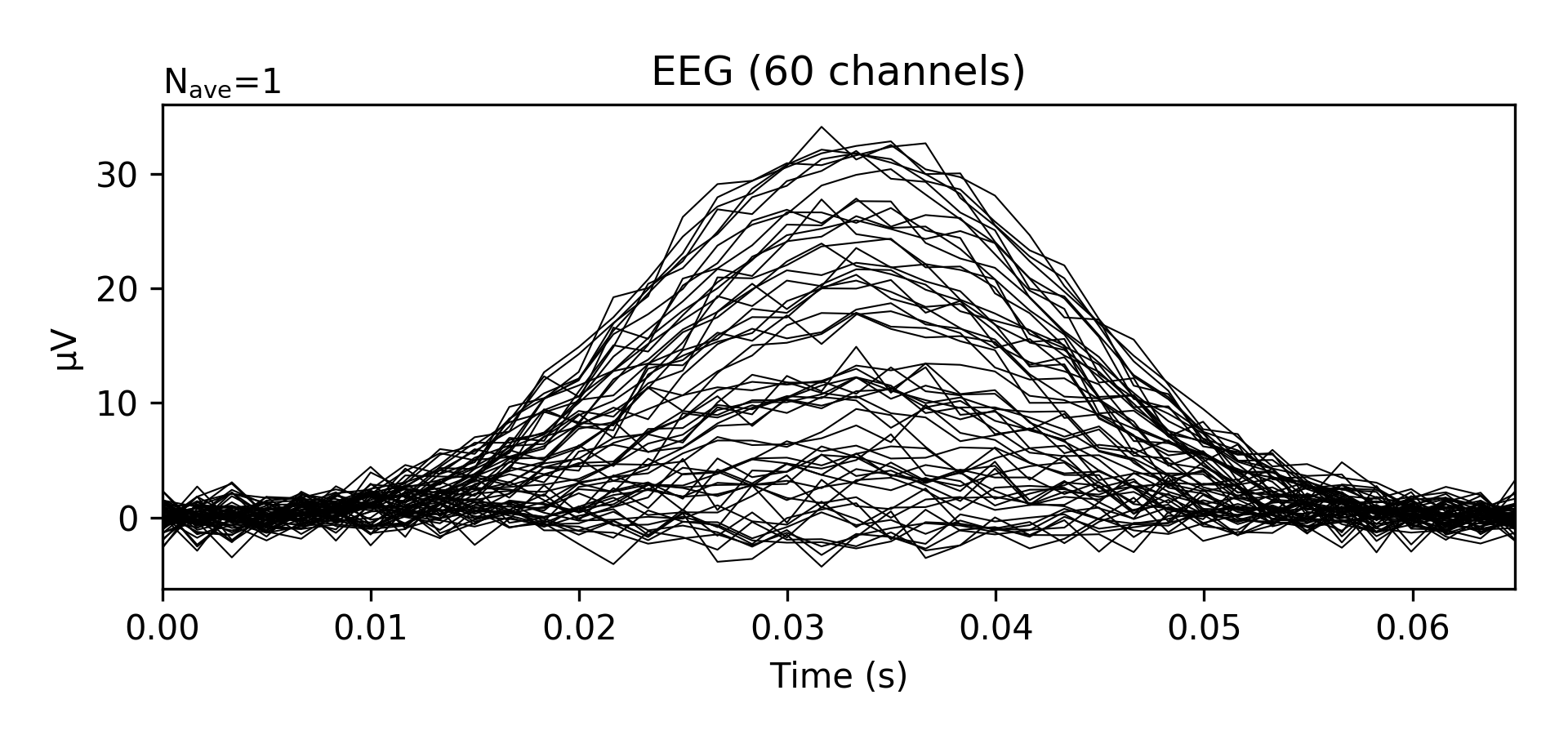}}
\caption{An example of simulated MEG (left) and EEG (right) recordings.}
\label{fig:data_fig}
\end{figure}
An example of simulated MEG and EEG recordings is pictured in Figure \ref{fig:data_fig}.

\medskip
We now perform source modeling from these data using both h-SESAME and SESAME. 
To avoid inverse crime \cite{kaipio2007statistical}, we first define, in the same way as before, a new brain discretization which is different from that used to generate the data, 
and then compute the corresponding MEG and EEG forward operators. This new brain grid is coarser and contains $7498$ points. 
From each dataset we extract the $20$ topographies centered around the signal peak. We set the number of particles to $100$ for both h-SESAME and SESAME.
As for the noise covariance matrix, we assume $\Gamma_N$ to be proportional to the identity matrix and we estimate the constant of proportionality separately
for each dataset, as the $20\%$ of the maximum signal value.
To asses how strongly the posterior distribution given by h-SESAME depends on the possibly inaccurate value of the hyperparameter $\sigma_{min}$, and to compare 
it to the dependence of SESAME result on $\sigma_q$, we analyze each dataset with both h-SESAME and SESAME in three different algorithm settings. 
To this aim, we introduce a scale factor $k \in  \{0.1,1,10\}$, henceforth called the prior scale factor, and then we set 
$\hat{\sigma}_q\/(k)\, =\, 2\@k\@\cdot\@10^{-7}$, $\hat{\sigma}_{min}\/(k) = \frac{\hat{\sigma}_q\/(k)}{35}$.
In this way, for $k$ in ascending order, we are respectively under-estimating, rightly-estimating and over-estimating the 
current intensity. All considered, we thus perform a total of $2400$ analyses,
$1200$ with h-SESAME and $1200$ with SESAME.

\medskip
From the approximated posterior distributions, the following quantities are then computed for subsequent performance evaluation:

\begin{itemize}
\item estimated number of sources $\hat{n}_D$, as the argmax of the marginal posterior distribution of the number of dipoles;
\item estimates of ECD locations as the local peaks of the probability map:
\begin{equation}
\label{eq:pmap}
\mathbb{P}\/(r\@|\@\mbf{y}_{1:T},\hat{n}_D) = \sum_{p=1}^{n_P} w^{\@p}\ \delta\left(\hat{n}_D^{},\, (n_D)^{\@p}\right) 
\sum_{d=1}^{(n_D)^{\@p}} \delta\left(r\, ,\@ (\@r^{\@d}\@)^{\@p}\right)\, .
\end{equation}
\end{itemize}

\subsection{Performance evaluation}

In order to quantitatively assess the differences between the two methods, we consider the following quantities. 

First, we construct confusion matrices for the estimated number of dipoles: here the row index represents the true number of dipoles (TND), the
column index represents the estimated number of dipoles (END), and each matrix element counts how many times that particular combination TND-END appears. 
The best result here would be to have a perfectly diagonal matrix, with all entries equal to 100 (because we have 100 datasets for each TND).

Then we compute a measure of the distance between the estimated dipole locations $\hat{r}^{\@1:\hat{n}_D}$ and the true dipole locations 
$r^{\@1:n_D}$; since these sets might have different cardinalities, and because we do not want to compute the distance of two different estimated dipoles 
from the same true dipole, we use a particular version of the Optimal Subpattern Assignment (OSPA) metric \cite{schuhmacher2008consistent}, defined as follows:

\begin{equation}
    OSPA = \min_{\phi} \sum_{d=1}^{\min(\hat{n}_D , n_D)} \bigg\|\, \hat{r}^{\@d} - r^{\@\phi(d)}\, \bigg\|\ ,
    \label{eq:OSPA}
\end{equation}
where $\phi$ represent all possible permutations.

We then proceed to assess how sensitive the posterior probability map in (\ref{eq:pmap}) is to the prior information on source strength\@: for each dataset, 
we calculate the three maps, corresponding to the three different algorithm settings, and then integrate over the brain volume the pairwise square differences:
\begin{equation}
    post\_var = \sum_{i\ne j} \int \bigg(\,\mathbb{P}\/(r\@|\@\mbf{y}_{1:T},\hat{n}_D, \sigma_i) - \mathbb{P}\/(r\@|\@\mbf{y}_{1:T},\hat{n}_D, \sigma_j)\,\bigg)^2\ d\/r\, ,
    \label{eq:post_var}
\end{equation}
where $\sigma_i$ refers to $\hat{\sigma}_{min}\/(i)$ for h-SESAME and to $\hat{\sigma}_q\/(i)$ for SESAME.

Finally, we check whether the final value of the hyperparameter $\sigma_q$ estimated by h-SESAME is anywhere close to the true source intensity.

\subsection{Results}

Whenever applicable, we report the results in the Figures using the following scheme: the results obtained from EEG data are given in the top panel, 
while the results obtained from MEG data are shown in the bottom panel; 
in each panel, the top row corresponds to SESAME results (in red), and the bottom row to h-SESAME results (in blue); then from left to right each column correspond to a different value of the prior scale factor $k$, in ascending order. 

\begin{figure}[ht]
    \centering    
    \subfloat[EEG results]{\includegraphics[width=\textwidth]{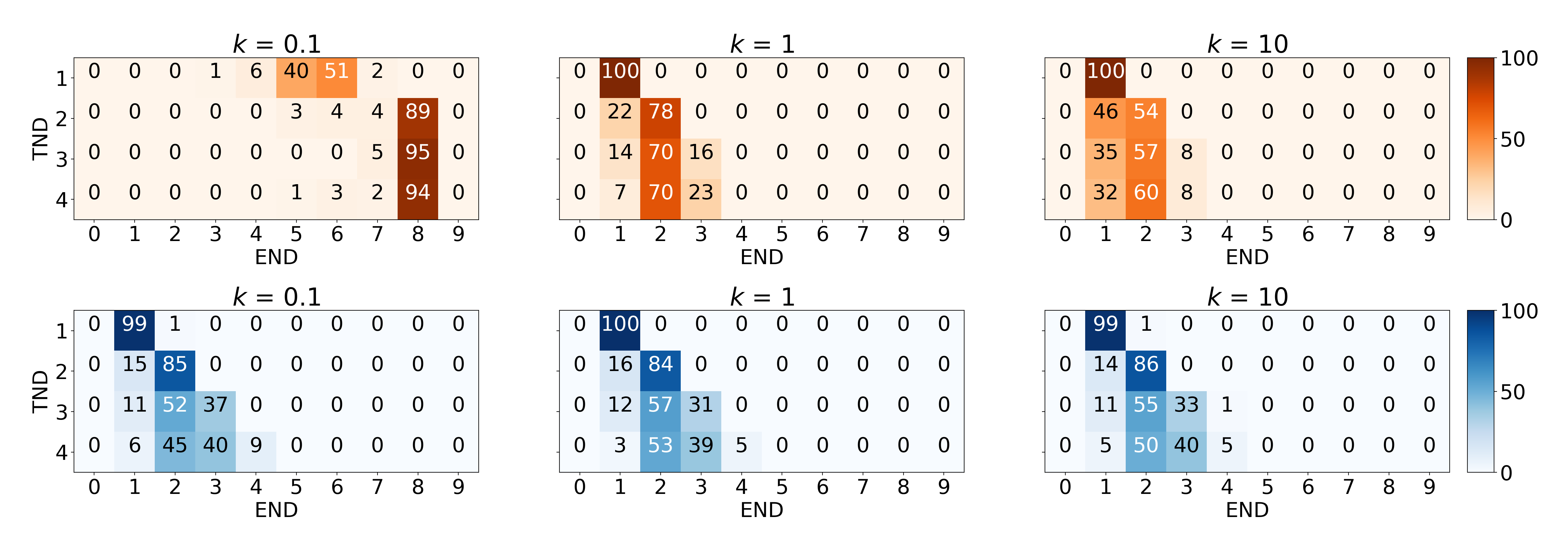}}\\
    \subfloat[MEG results]{\includegraphics[width=\textwidth]{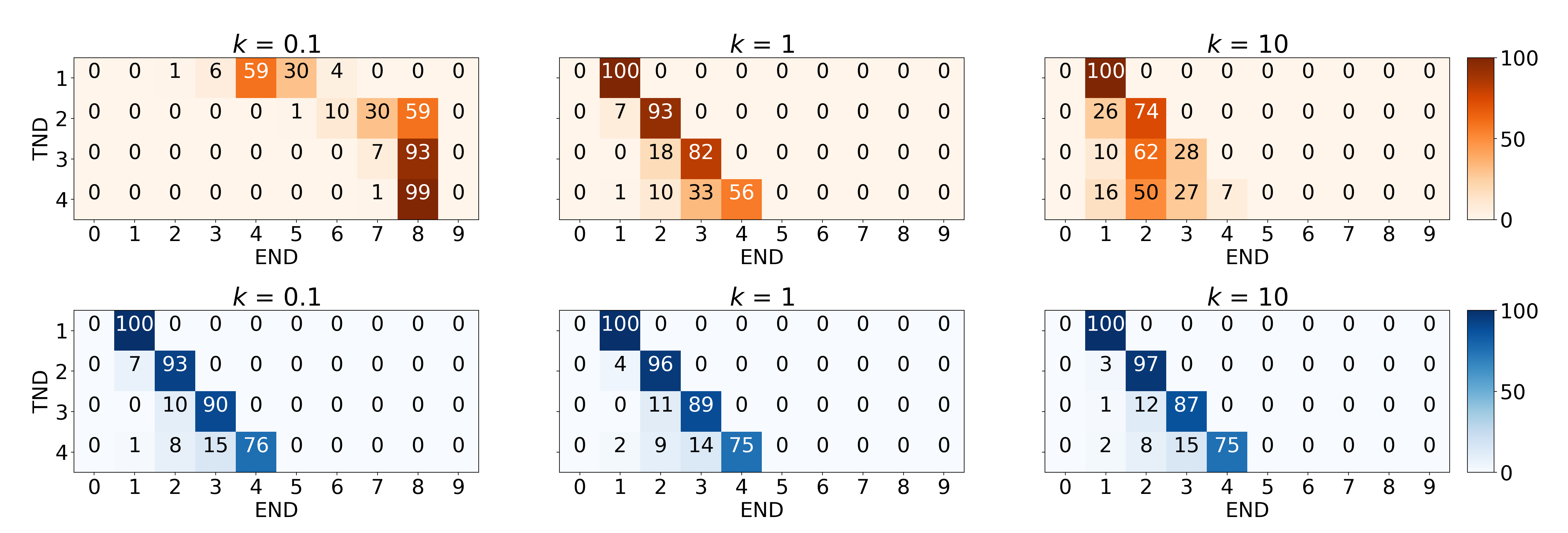}}
    \caption{Confusion matrices for the estimated number of dipoles, for three different values of the prior scale factor $k$: in the top panel results obtained with simulated EEG data, in the bottom panel results obtained with simulated MEG data.}
    \label{fig:CM}
\end{figure}

In Figure \ref{fig:CM} we report the confusion matrices on the estimated number of sources. For both EEG and MEG, SESAME tends to a strong overestimation of the number of dipoles when the prior scale factor $k$ is set to $0.1$, to a slight underestimation with $k = 1$, and to a stronger understimation with $k=10$. The END of h-SESAME, on the other hand, does not seem to depend on the value of the hyperparameter, and always provides better confusion matrices than SESAME; we notice that a slight understimation is expected due to the Poisson prior that encourages parsimonious models.

\begin{figure}[ht]
    \centering
    \subfloat[EEG results]{\includegraphics[width=\textwidth]{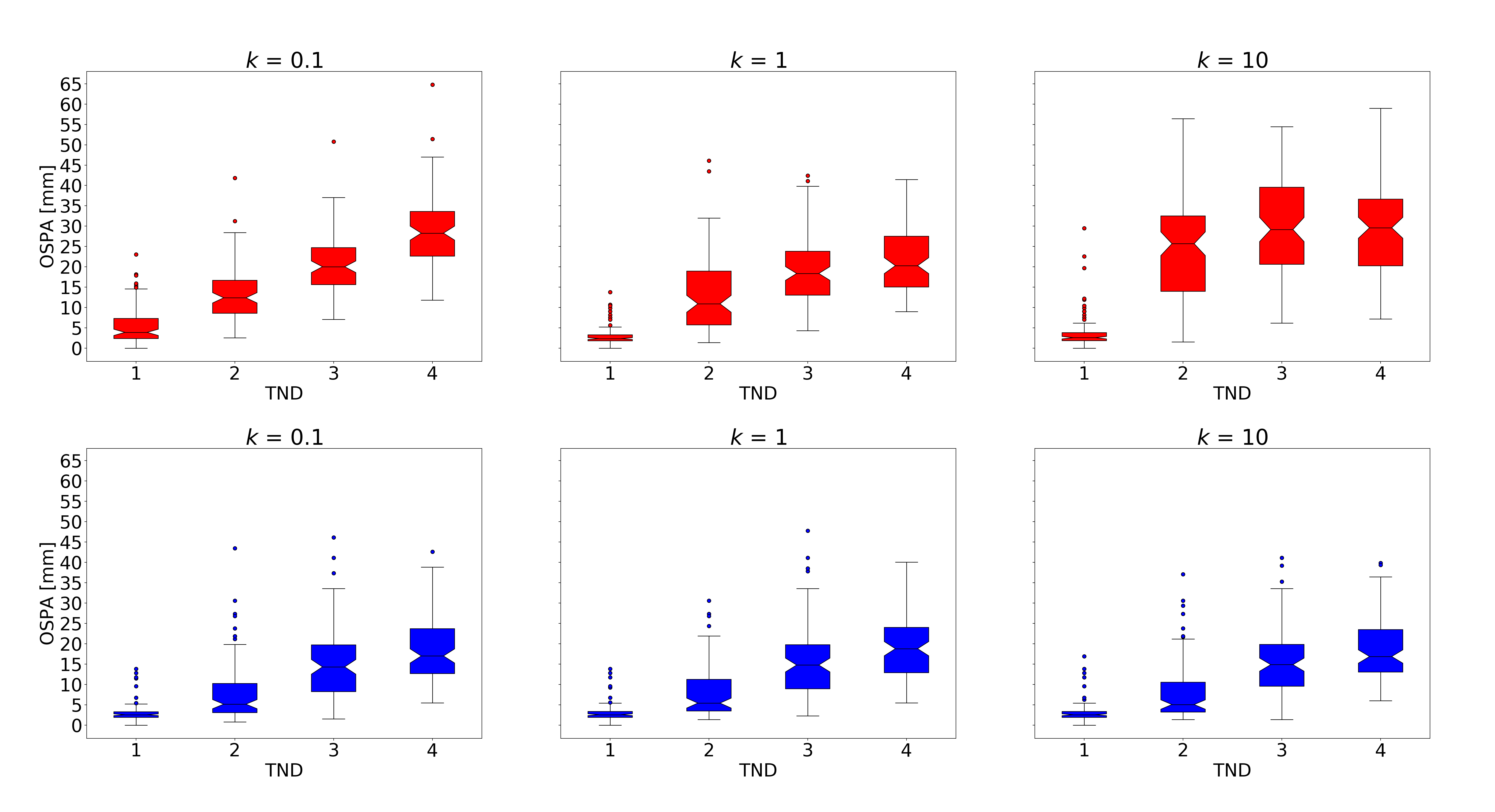}}\\
    \subfloat[MEG results]{\includegraphics[width=\textwidth]{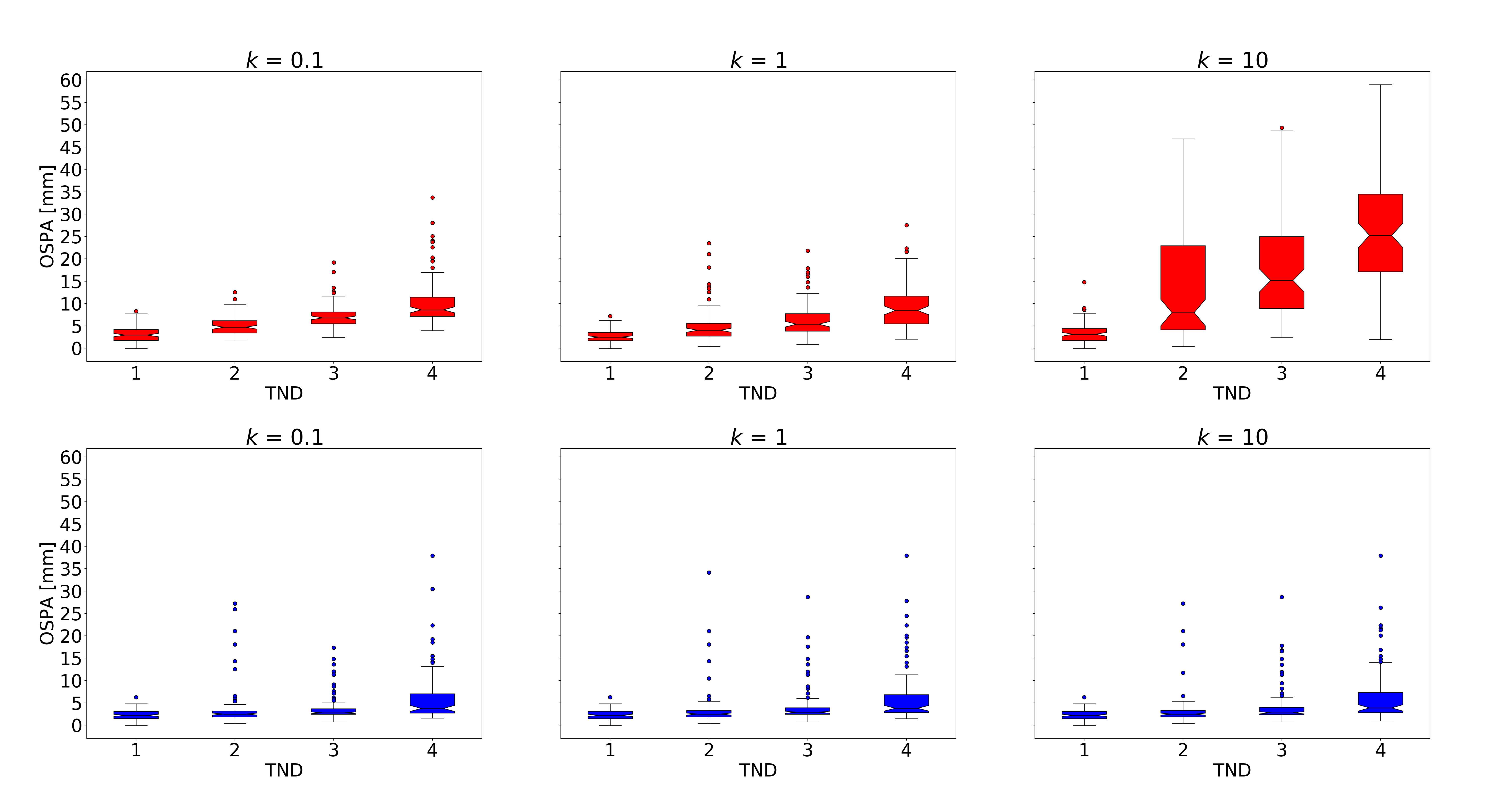}}
    \caption{Boxplots of the OSPA metric, quantifying the distance between the true and estimated dipole configurations, for three different values of the prior scale factor $k$: in red the SESAME results, in blue the h-SESAME results; in the top panel results obtained with simulated EEG data, in the bottom panel results obtained with simulated MEG data.}
    \label{fig:OSPA}
\end{figure}

In Figure \ref{fig:OSPA} we report the boxplot of the OSPA metric, assessing the distance between estimated and true ECDs. 
We observe that, as expected, there is a growth of the OSPA with the complexity of the source configuration. 

For SESAME, the worst case is always the case when 
$k = 10$\@: here, the Gaussian prior $\pi(\@\mbf{q}^{\@d}_{\@1:T}\/)$ possibly forces too intense currents that are therefore biased towards deeper locations in the brain. 
On the other hand,  when $k=0.1$, a larger number of ECDs is estimated by SESAME, but the OSPA remains similar to the case $k=1$: this is because the location of estimated ECDs are similar to those of $k=1$, and the additional sources are not actually used in the calculation of the OSPA, which only takes into account the best $\min(\hat{n}_D,n_D)$ ECDs.

The h-SESAME algorithm performs systematically better than SESAME, partly as a consequence of a better END, as seen in the confusion matrices above. 
The improvement is particularly manifest in the case of MEG data, where the OSPA remains very low up to TND$\,=4$. 
The results of h-SESAME are, again, substantially independent on the value of the hyperparameter.


\begin{figure}[ht]
    \centering
    \subfloat[EEG results]{\includegraphics[width=\textwidth]{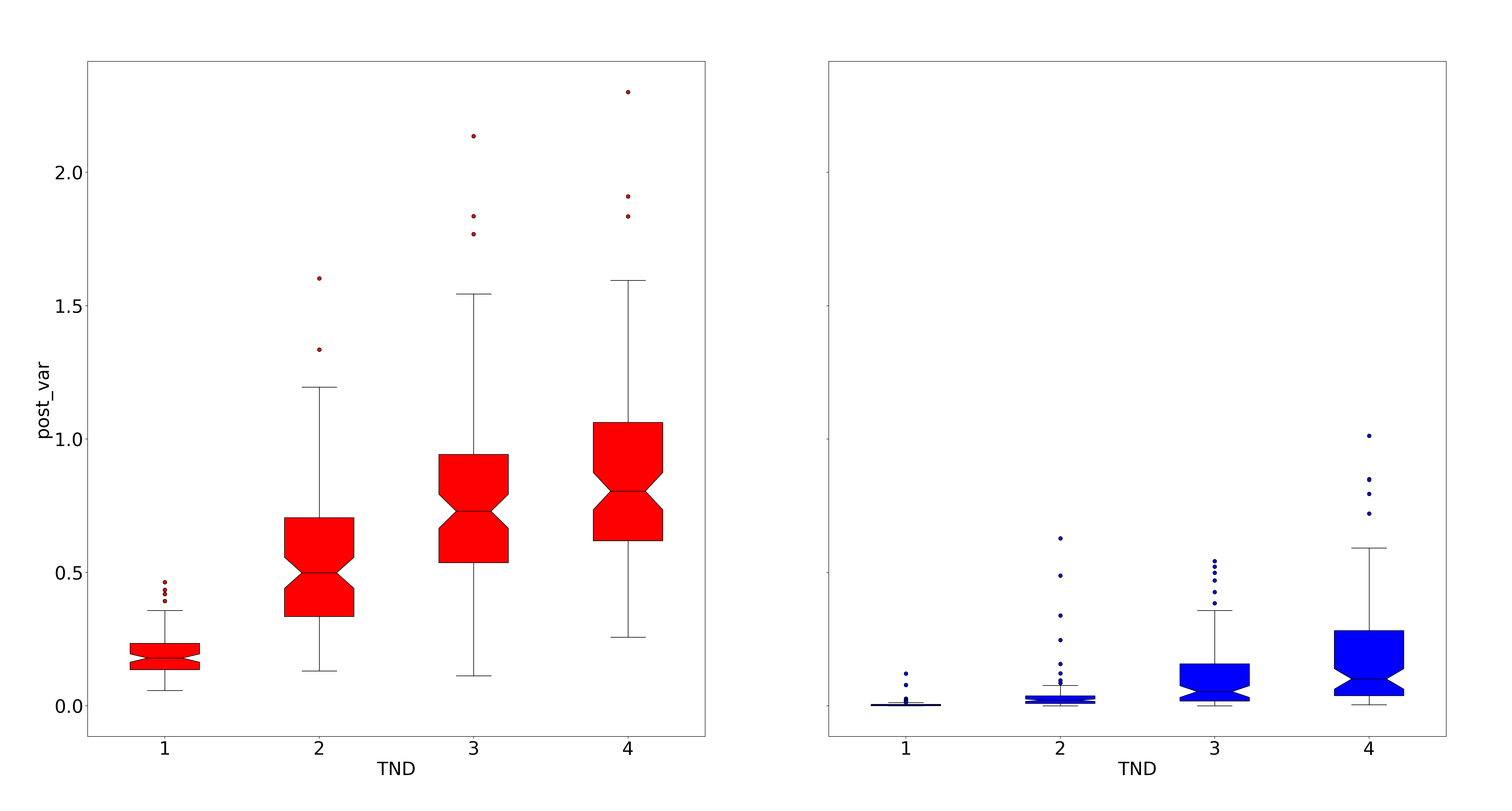}}\\
    \subfloat[MEG results]{\includegraphics[width=\textwidth]{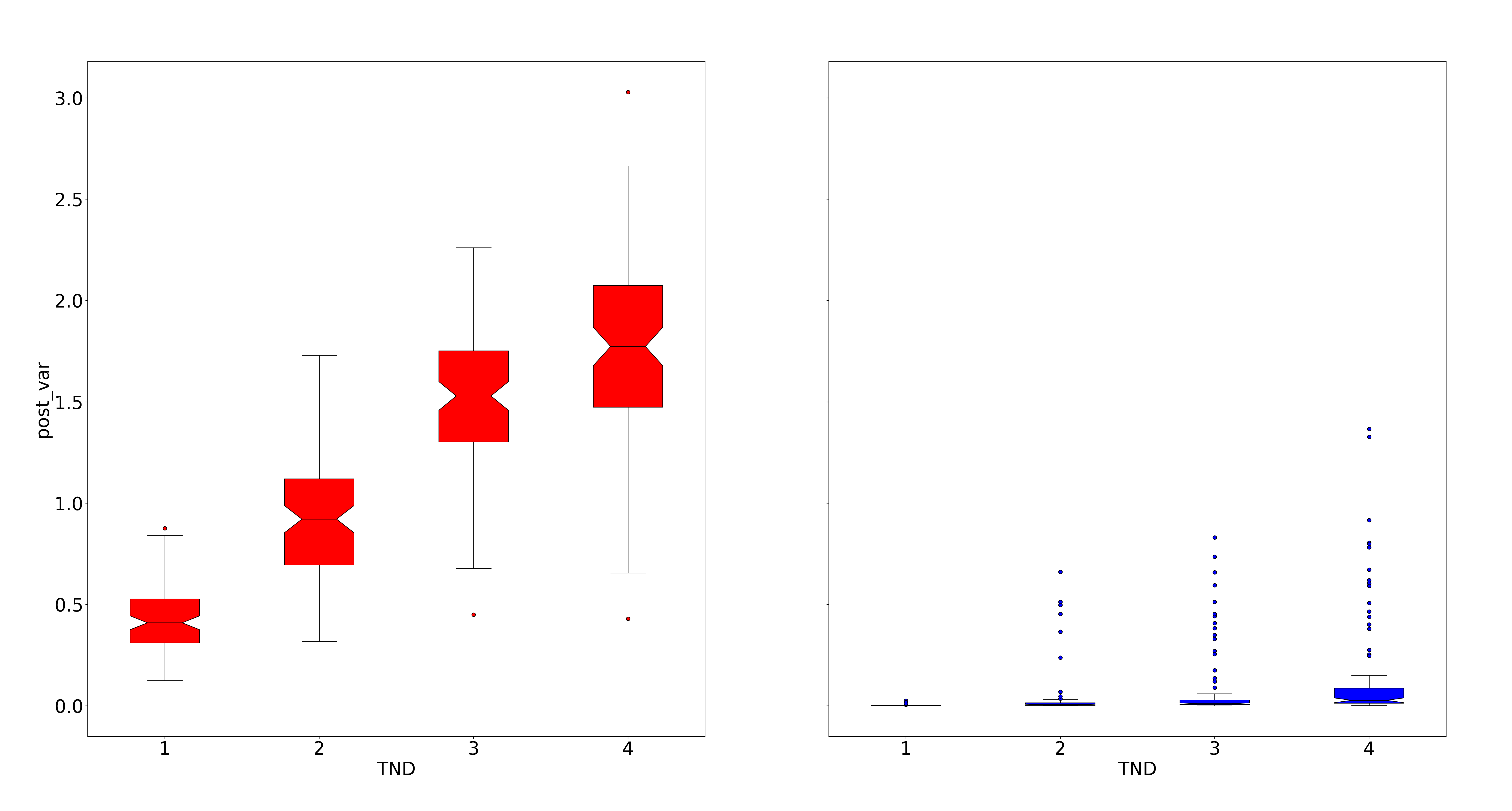}}
    \caption{Variance of the posterior probability map with respect to different values of the prior scale factor $k$, as defined in eq. (\ref{eq:post_var}): in red the SESAME results, in blue the h-SESAME results.}
    \label{fig:post_var}
\end{figure}

In Figure \ref{fig:post_var} we report the variance of the posterior distribution as described by equation (\ref{eq:post_var}). 
This result confirms that the posterior distributions corresponding to different values of the prior scale factor are substantially different for SESAME, and extremely similar for h-SESAME.

\begin{figure}
    \centering
    \subfloat[EEG results]{\hspace*{-20pt}\includegraphics[width=510pt]{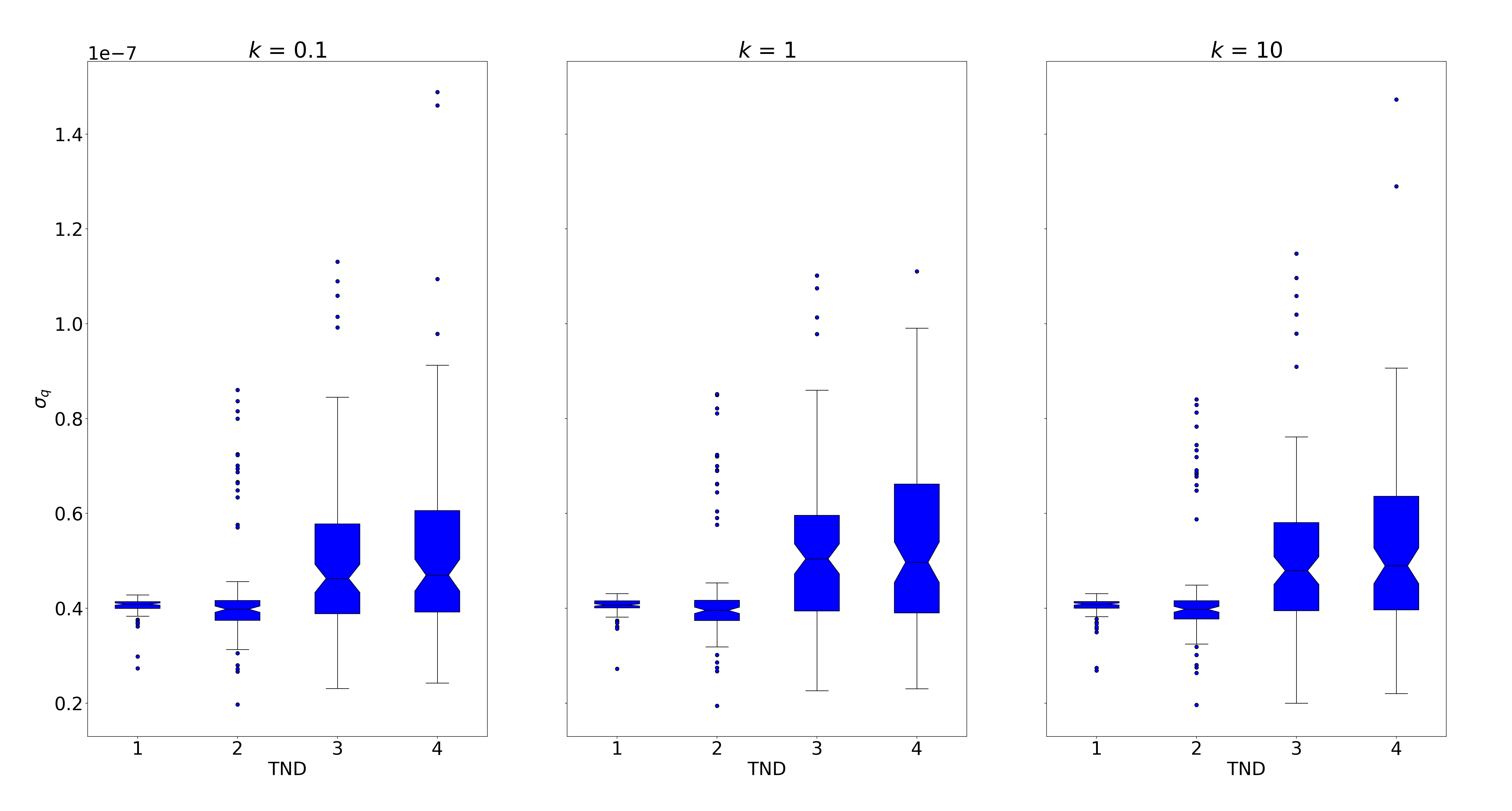}}\\
    \subfloat[MEG results]{\hspace*{-20pt}\includegraphics[width=510pt]{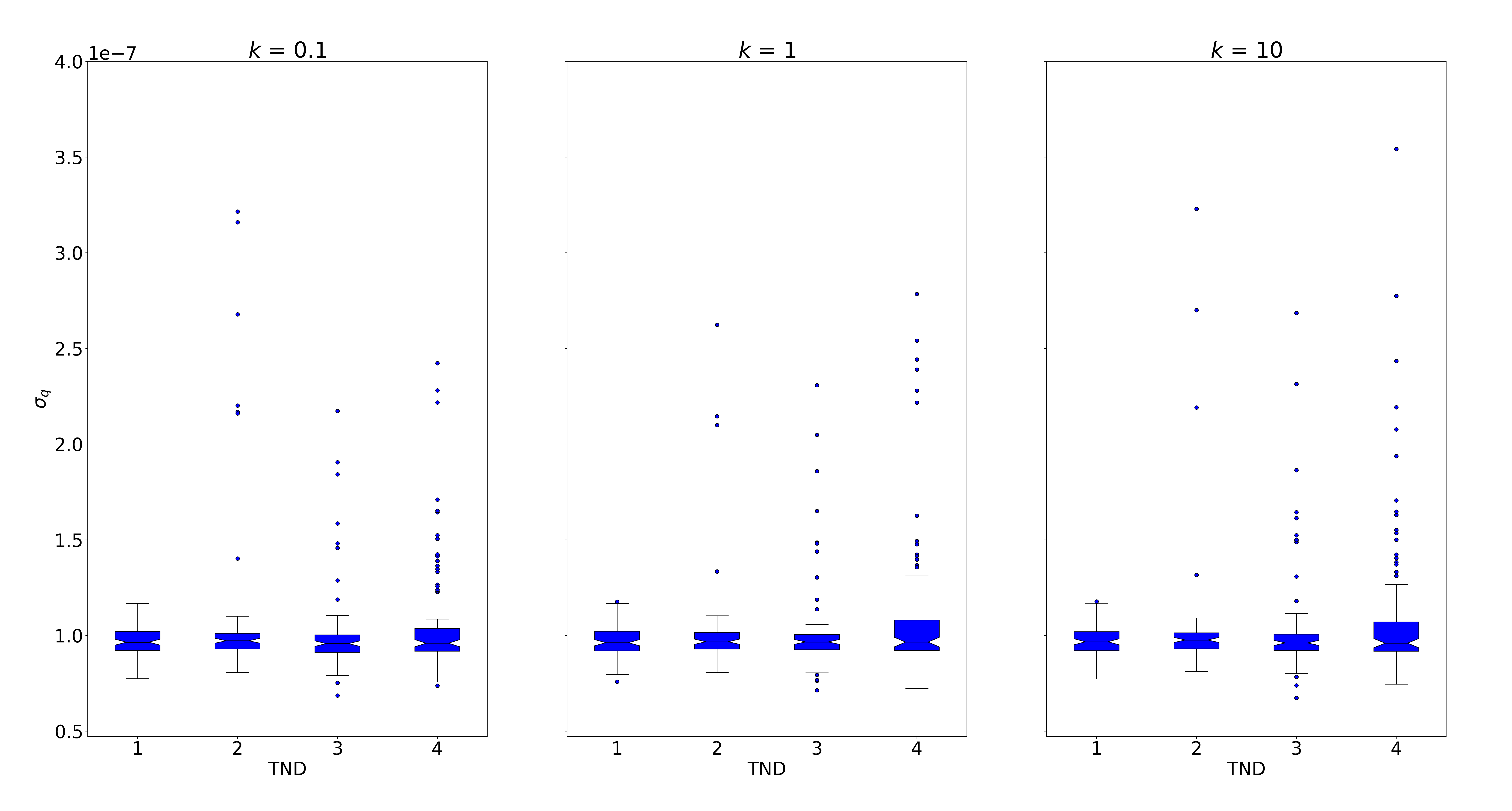}}
    \caption{Estimated value of the prior width $\sigma_q$, for different values of the prior scale factor $k$.}
    \label{fig:sq}
\end{figure}

Finally, in Figure \ref{fig:sq} we report the boxplots of the estimated value of the hyperparameter. We recall that, for simplicity, all simulations were performed with a maximum source strength equal to $1 e^{-7}$. The results are once again substantially independent on the value of the prior scale factor. We also notice that the estimated value of $\sigma_q$ is very stable in the case of MEG data, and substantially independent on the number of sources. In the case of EEG data the estimated valued is more variable, possibly due to the smaller number of sensors and the, possibly related, tendency to underestimate the number of dipoles (cfr. Fig. \ref{fig:CM}).
 
\section{Experimental data}

\begin{figure}
    \centering
    \includegraphics[scale=0.6]{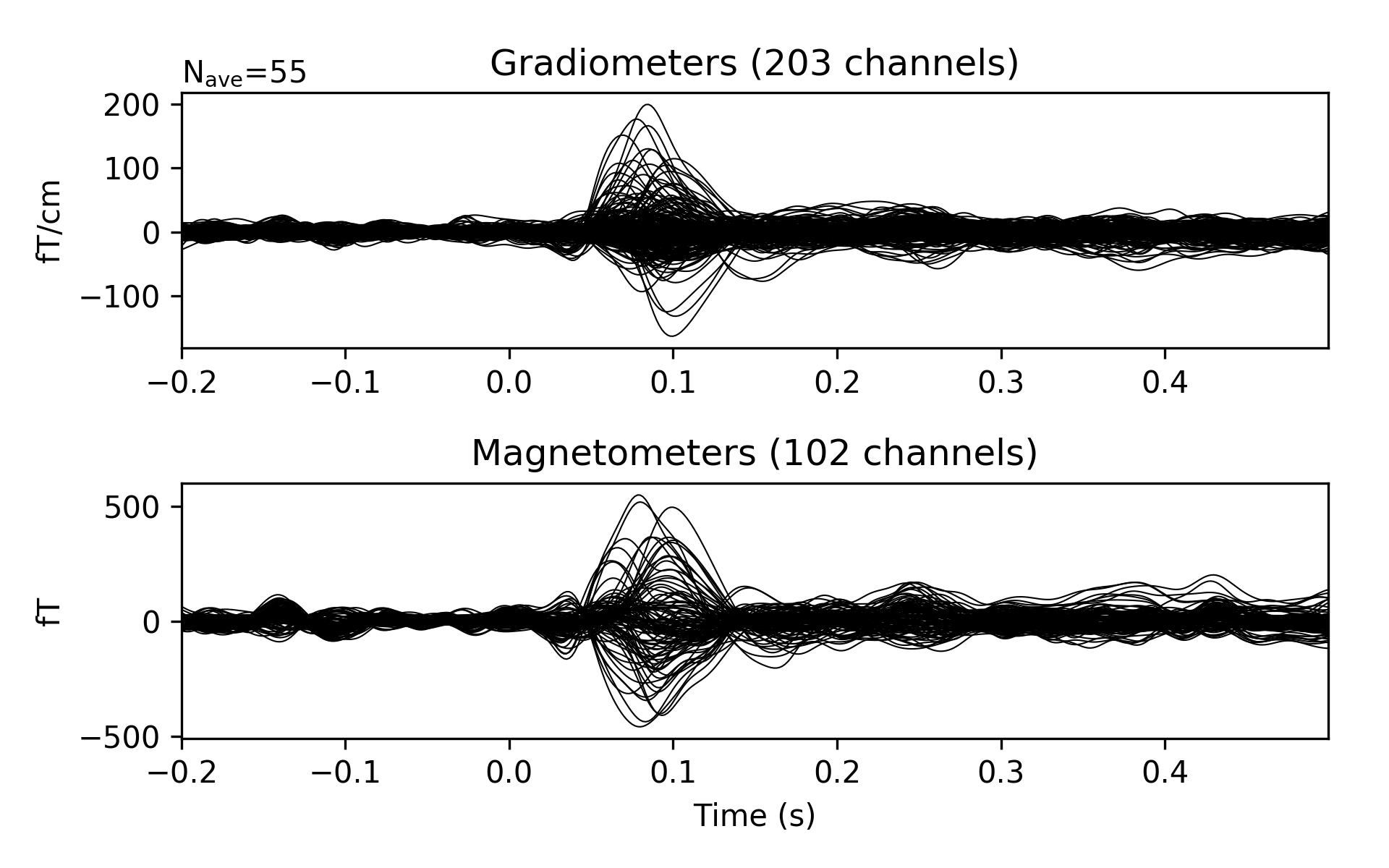}
    \caption{Experimental data.}
    \label{fig:exp_data}
\end{figure}

To further validate the performance of the proposed h-SESAME in a real scenario, we now show the results of source modeling from MEG data 
consisting in the average evoked response to auditory stimuli presented to the left ear. The analyzed data are once again 
taken from the \texttt{sample} open dataset which comes with the  MNE--Pyton package.
The description of the entire experiment can be found in \cite{gramfort2013meg, gramfort2013time} and is not repeated here.
Data are shown in Figure \ref{fig:exp_data} and correspond to $55$ averaged epochs.

We extract the topographies from $55\@$ms to $135\@$ms, we set again the number of particles to $100$ and we estimate the noise covariance as before.
Then we repeat the scheme already adopted in Section \ref{sec:num_sim}, namely we compare results from h-SESAME and SESAME in three different algorithm settings
corresponding to the different values of the prior scale factor\@: we set $\hat{\sigma}_q\/(k)\, =\, 3.2\@k\@\cdot\@10^{-8}$, 
$\hat{\sigma}_{min}\/(k) = \frac{\hat{\sigma}_q\/(k)}{35}$, where the value $3.2\@\cdot\@10^{-8}$ is estimated from the data and from the forward operator, and $k\in\{0.1,1,10\}$

\begin{figure}
\caption{Results of SESAME with $k=0.1$ (top row), $k=1$ (middle row) and $k=10$ (bottom row).}
\centering
\subfloat{\hspace*{-50pt}\includegraphics[scale=0.3]{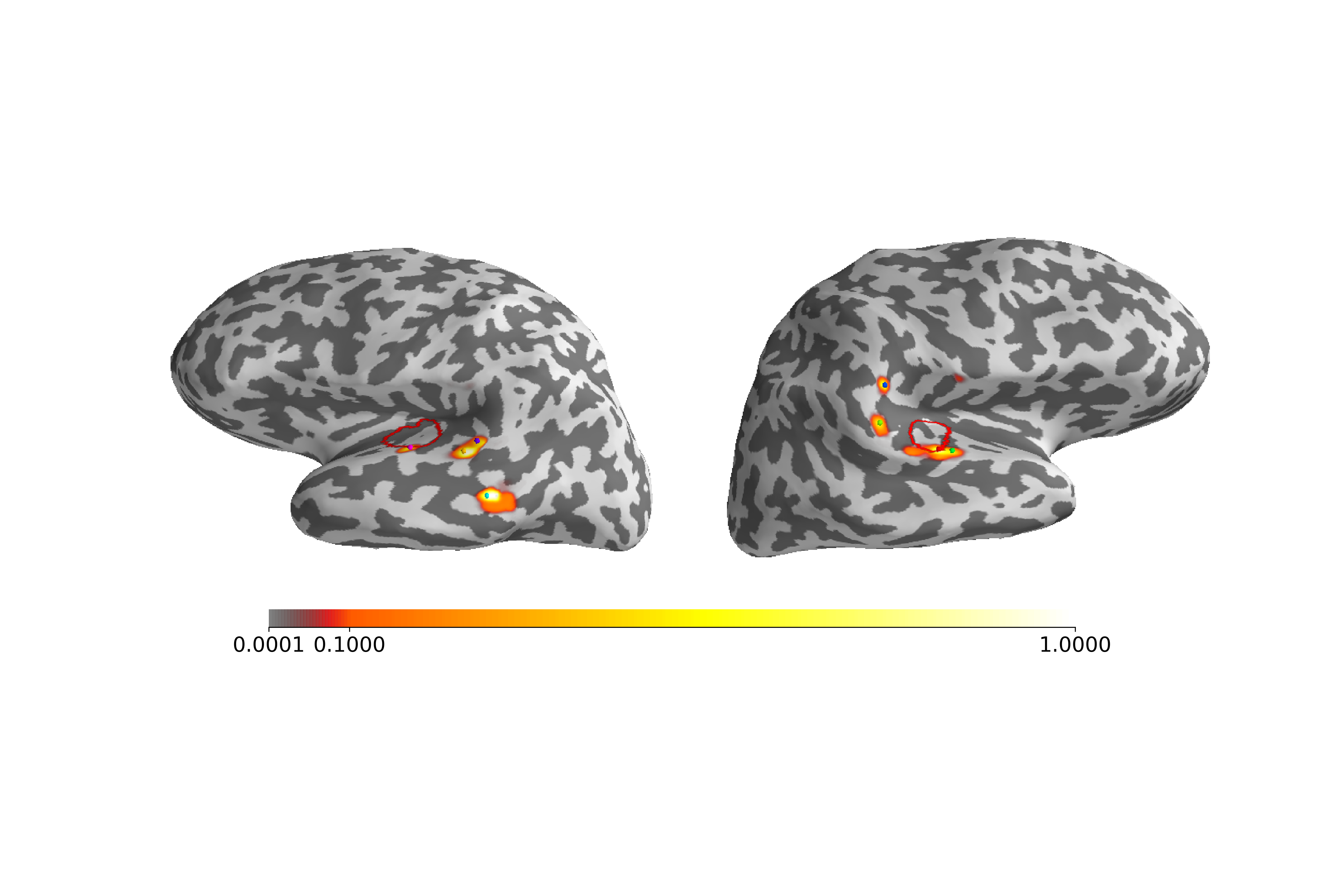}}
\hspace*{-20pt}
\subfloat{\includegraphics[scale=0.25]{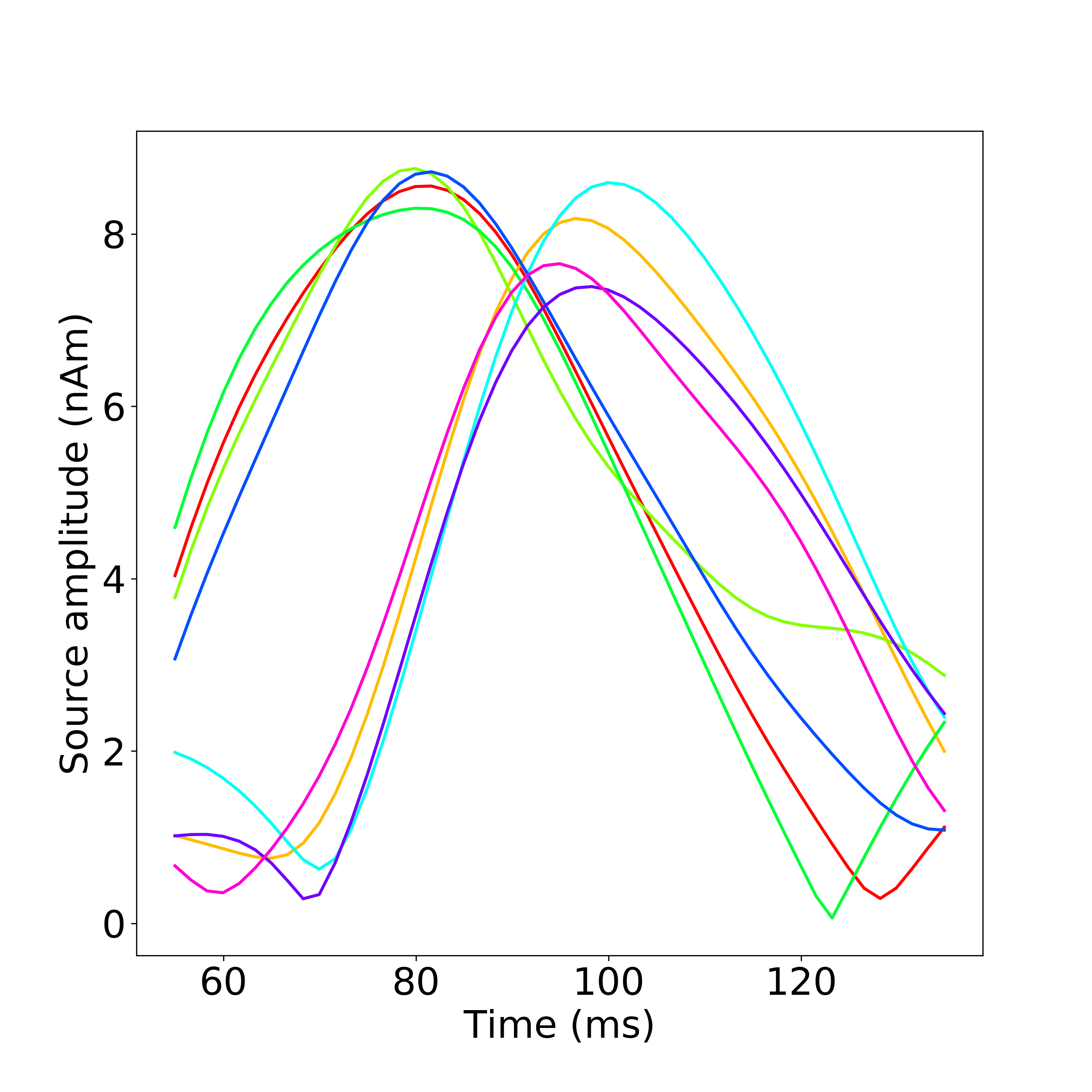}}\\[6pt]
\vspace*{-60pt}
\subfloat{\hspace*{-50pt}\includegraphics[scale=0.3]{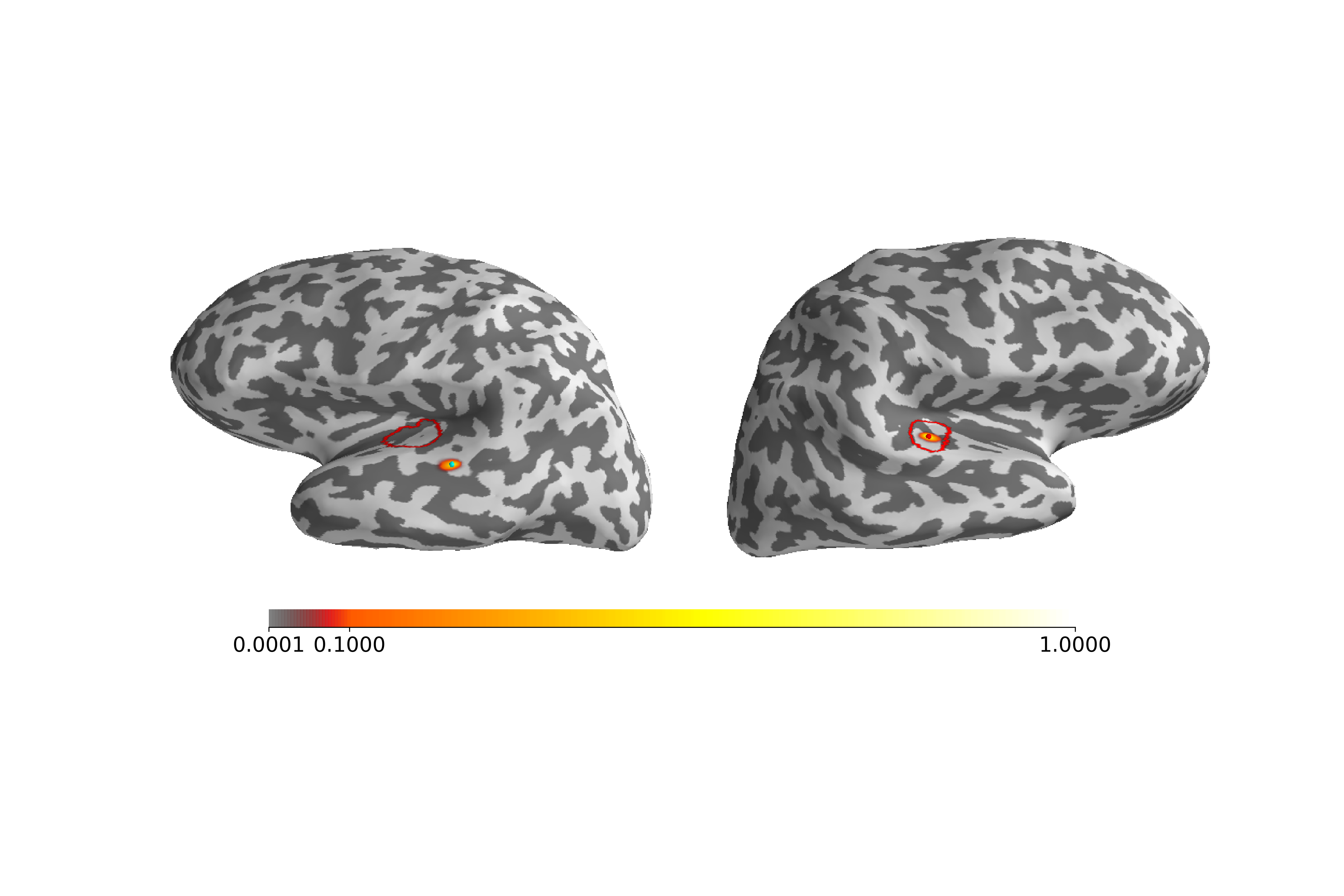}}
\hspace*{-20pt}
\subfloat{\includegraphics[scale=0.25]{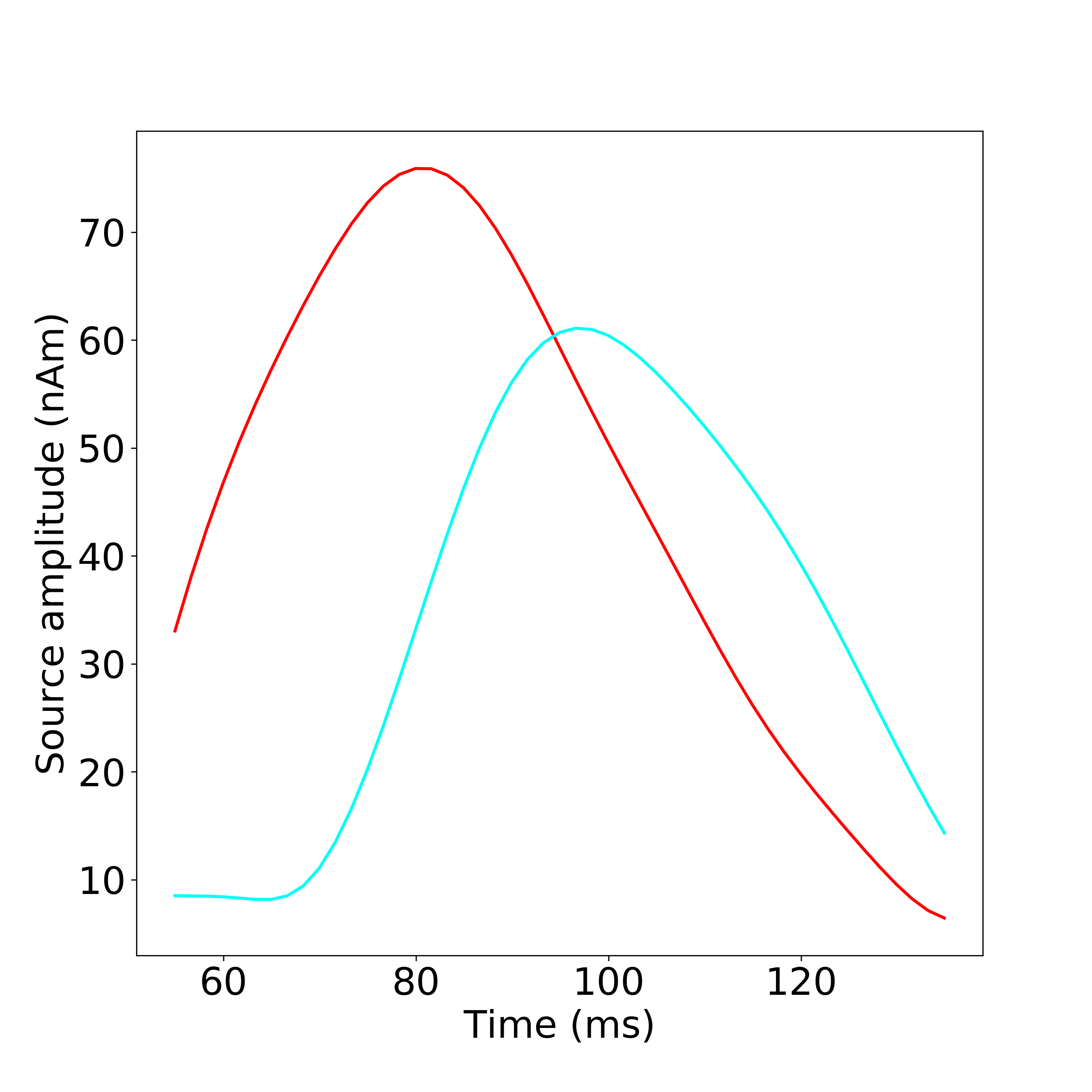}}\\[6pt]
\vspace*{-60pt}
\subfloat{\hspace*{-50pt}\includegraphics[scale=0.3]{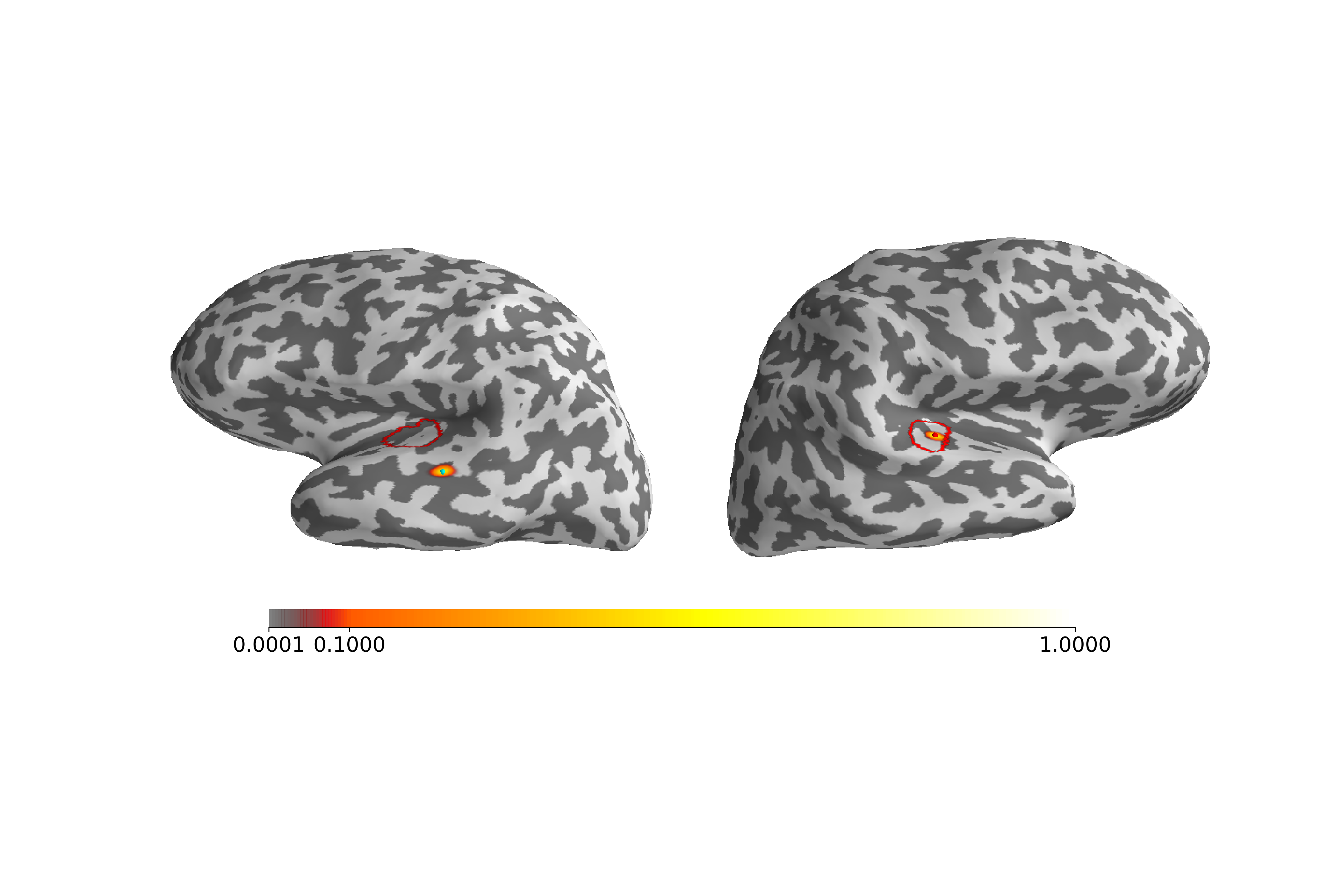}}
\hspace*{-20pt}
\subfloat{\includegraphics[scale=0.25]{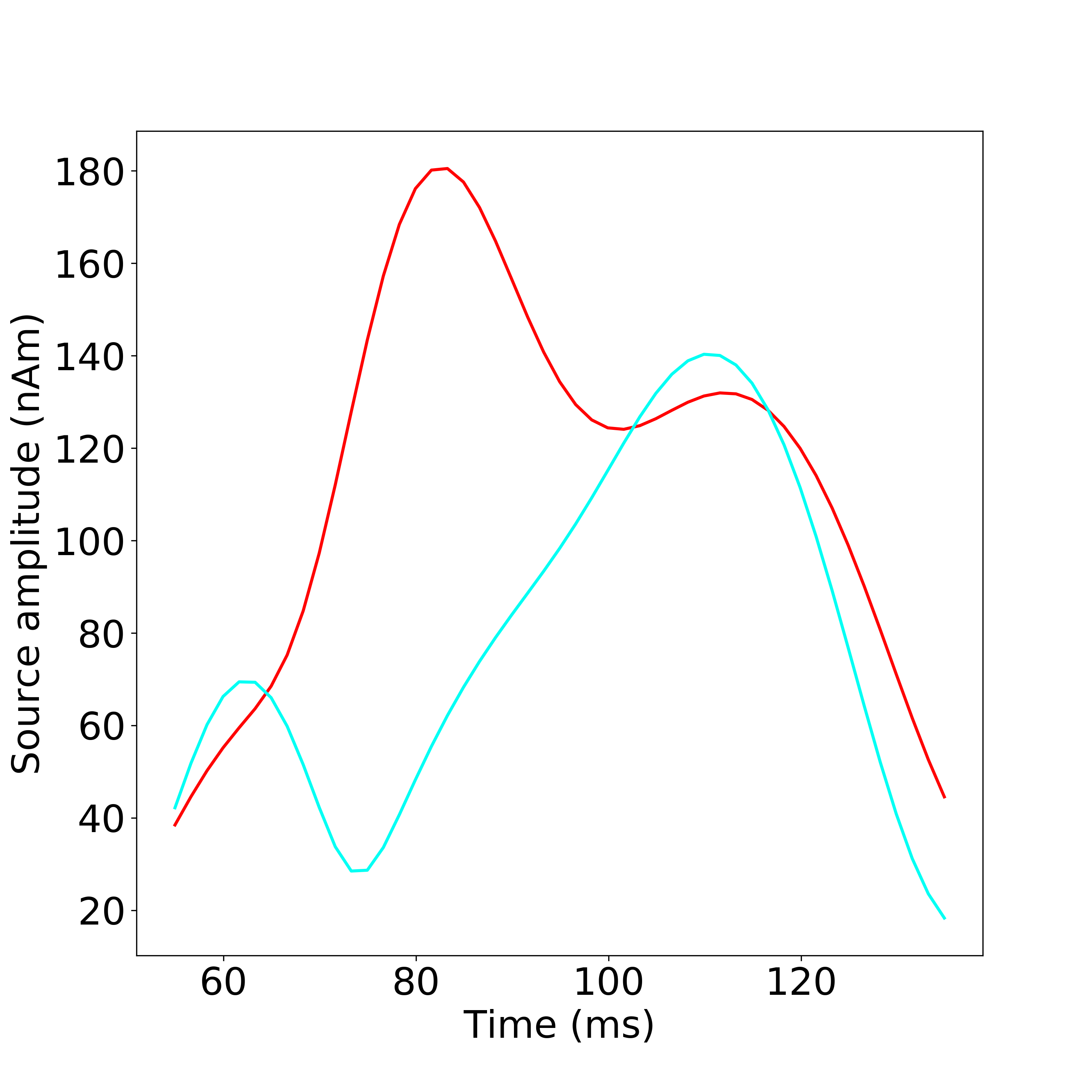}}
\label{fig:left_auditory_sesame_old}
\end{figure}

\begin{figure}
\caption{Results of h-SESAME with $k=0.1$ (top row), $k=1$ (middle row) and $k=10$ (bottom row).}
\centering
\subfloat{\hspace*{-50pt}\includegraphics[scale=0.3]{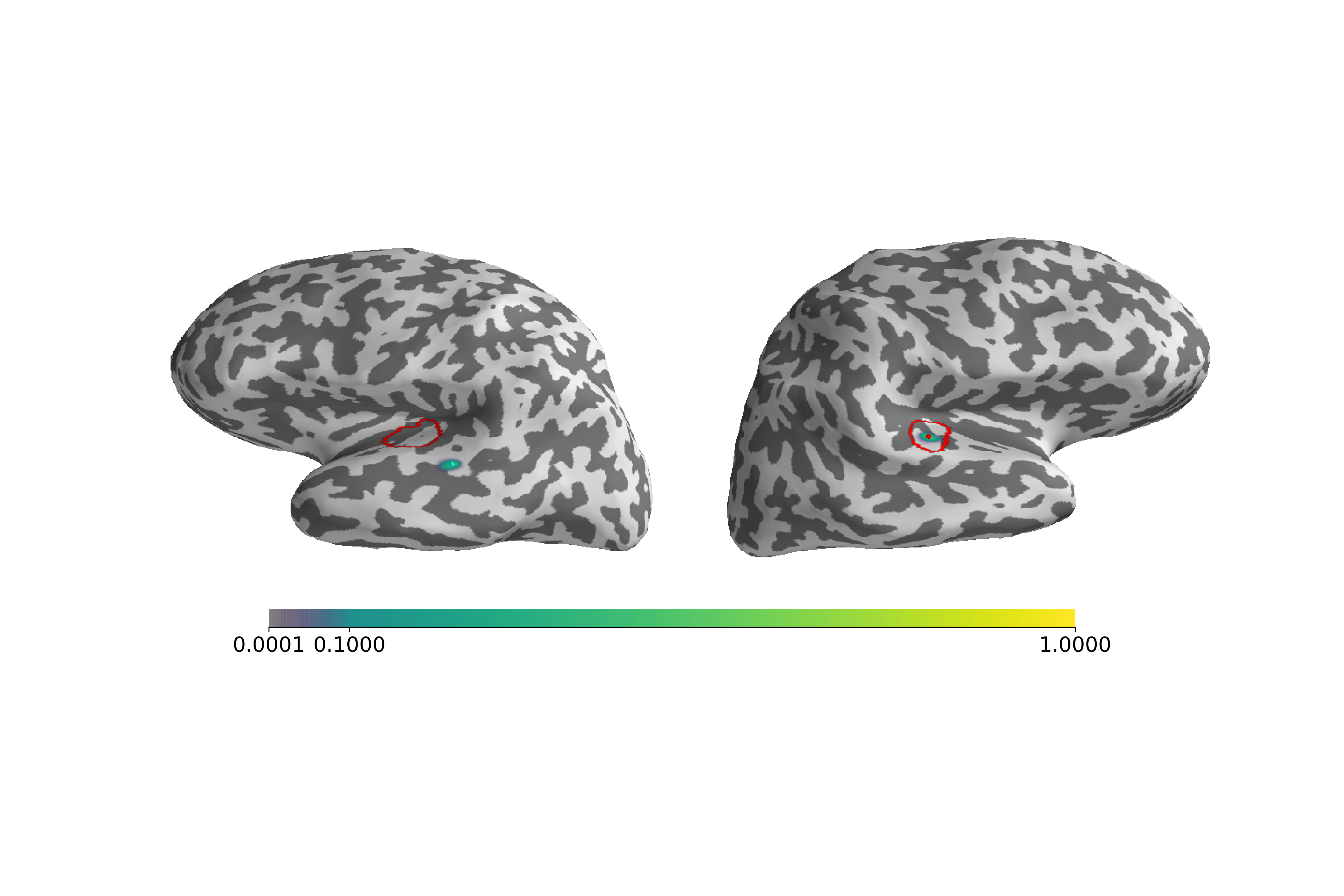}}
\hspace*{-20pt}
\subfloat{\includegraphics[scale=0.25]{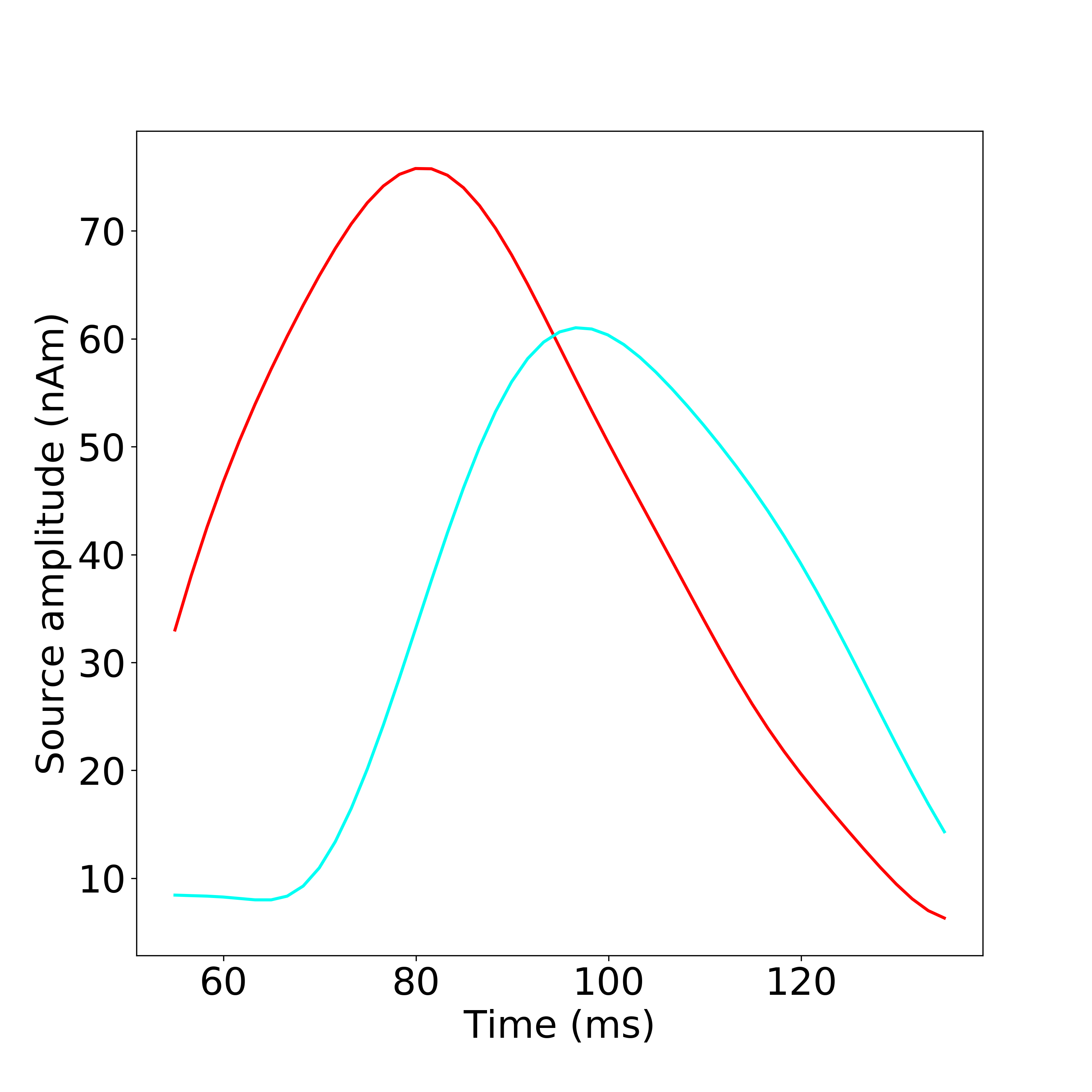}}\\[6pt]
\vspace*{-60pt}
\subfloat{\hspace*{-50pt}\includegraphics[scale=0.3]{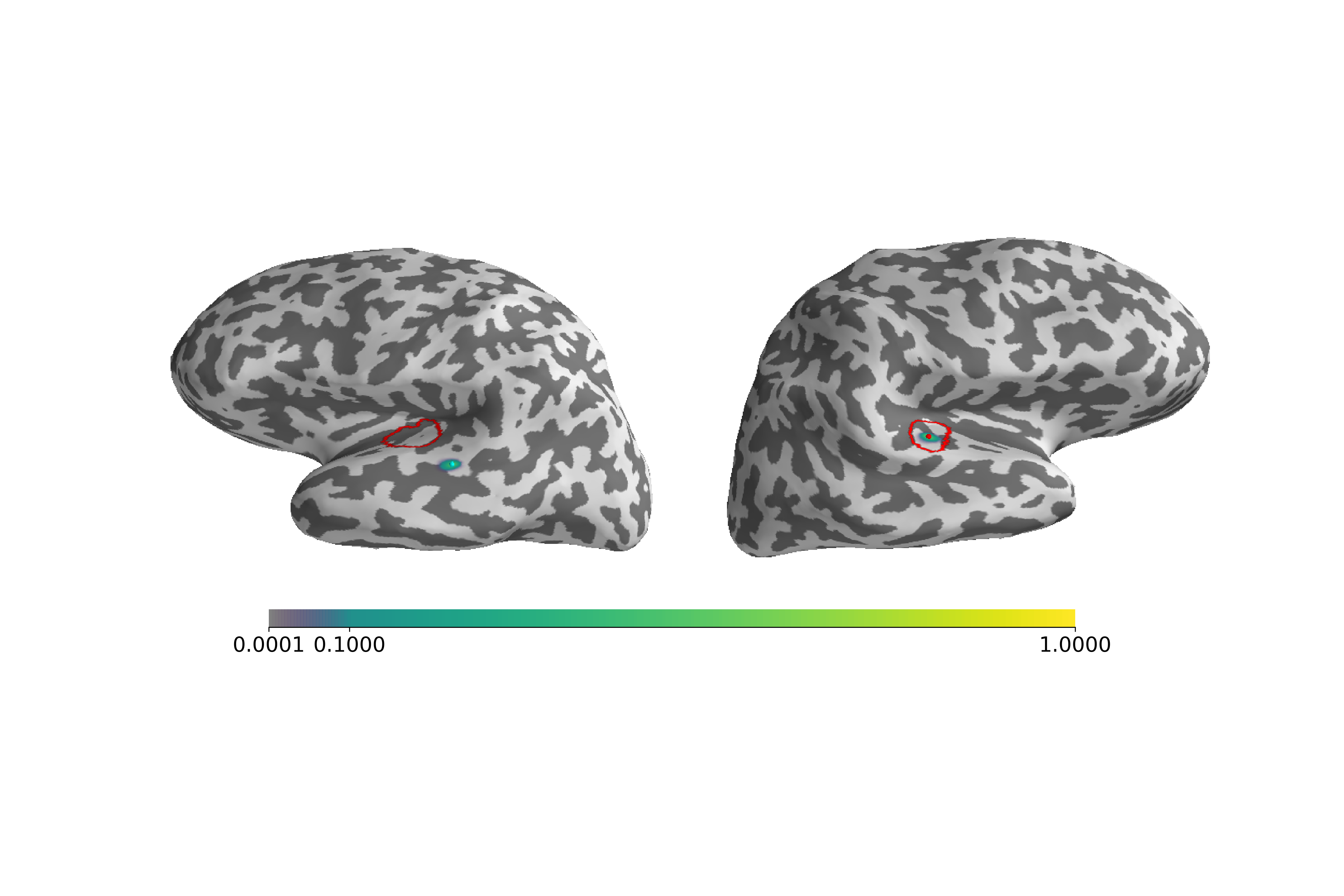}}
\hspace*{-20pt}
\subfloat{\includegraphics[scale=0.25]{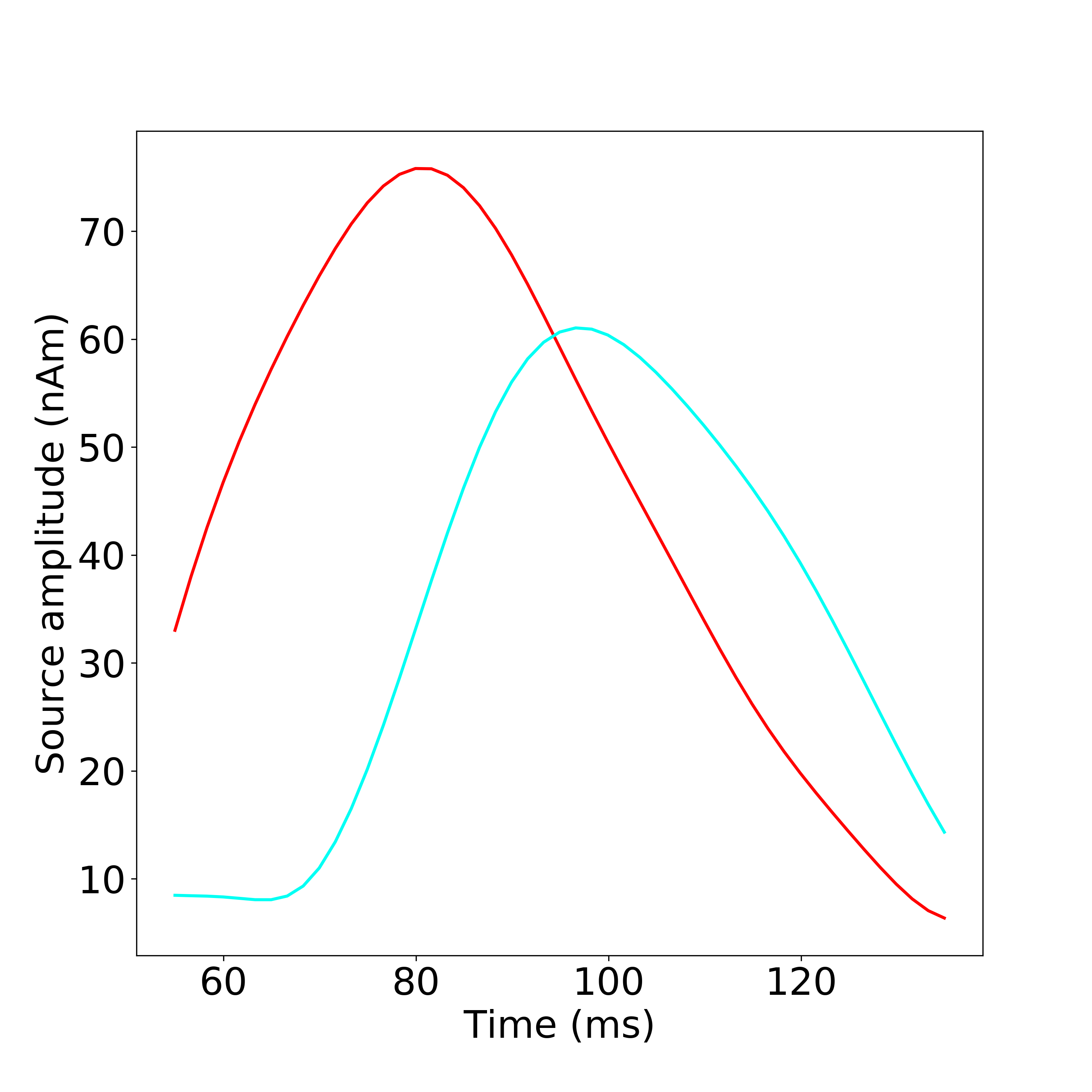}}\\[6pt]
\vspace*{-60pt}
\subfloat{\hspace*{-50pt}\includegraphics[scale=0.3]{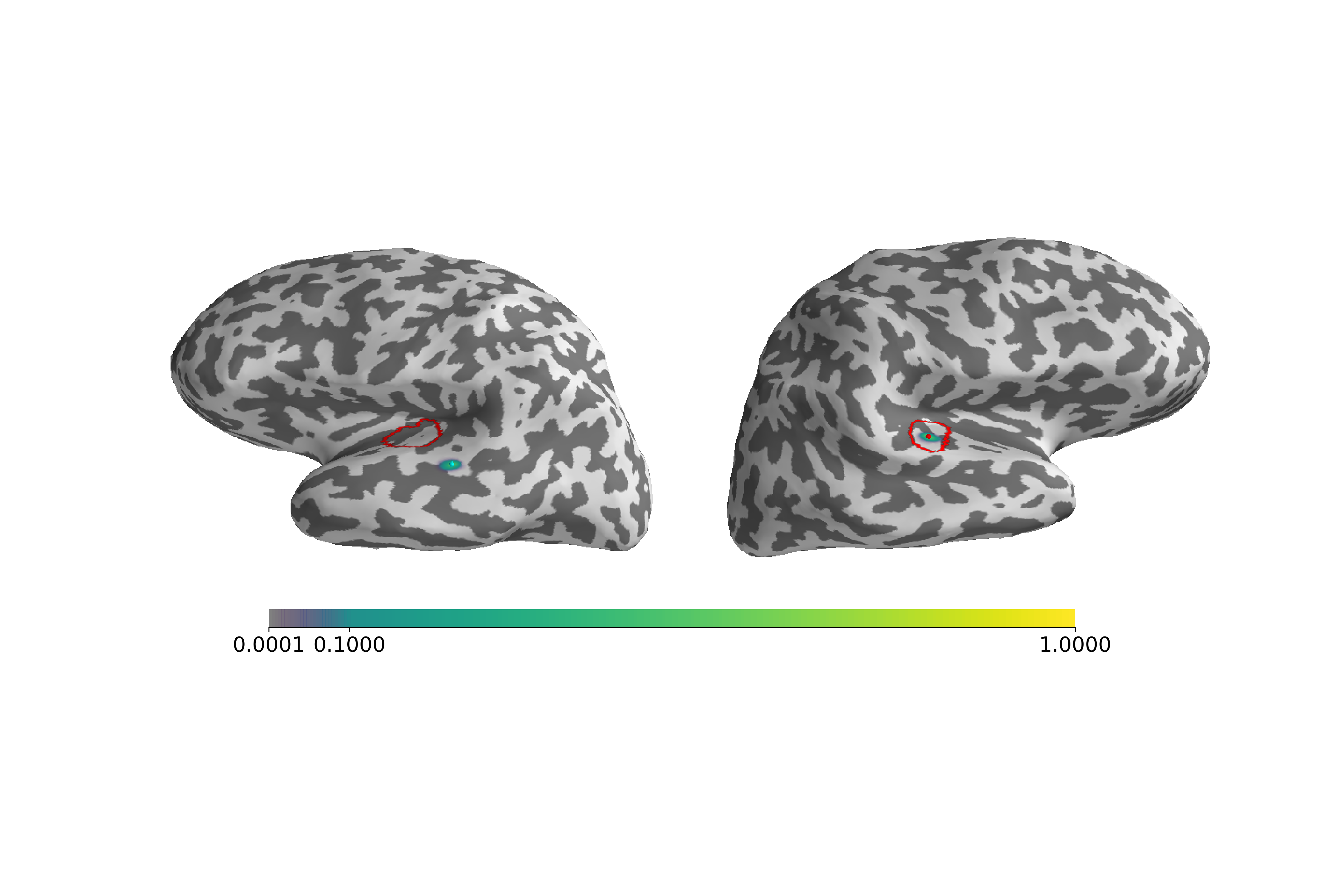}}
\hspace*{-20pt}
\subfloat{\includegraphics[scale=0.25]{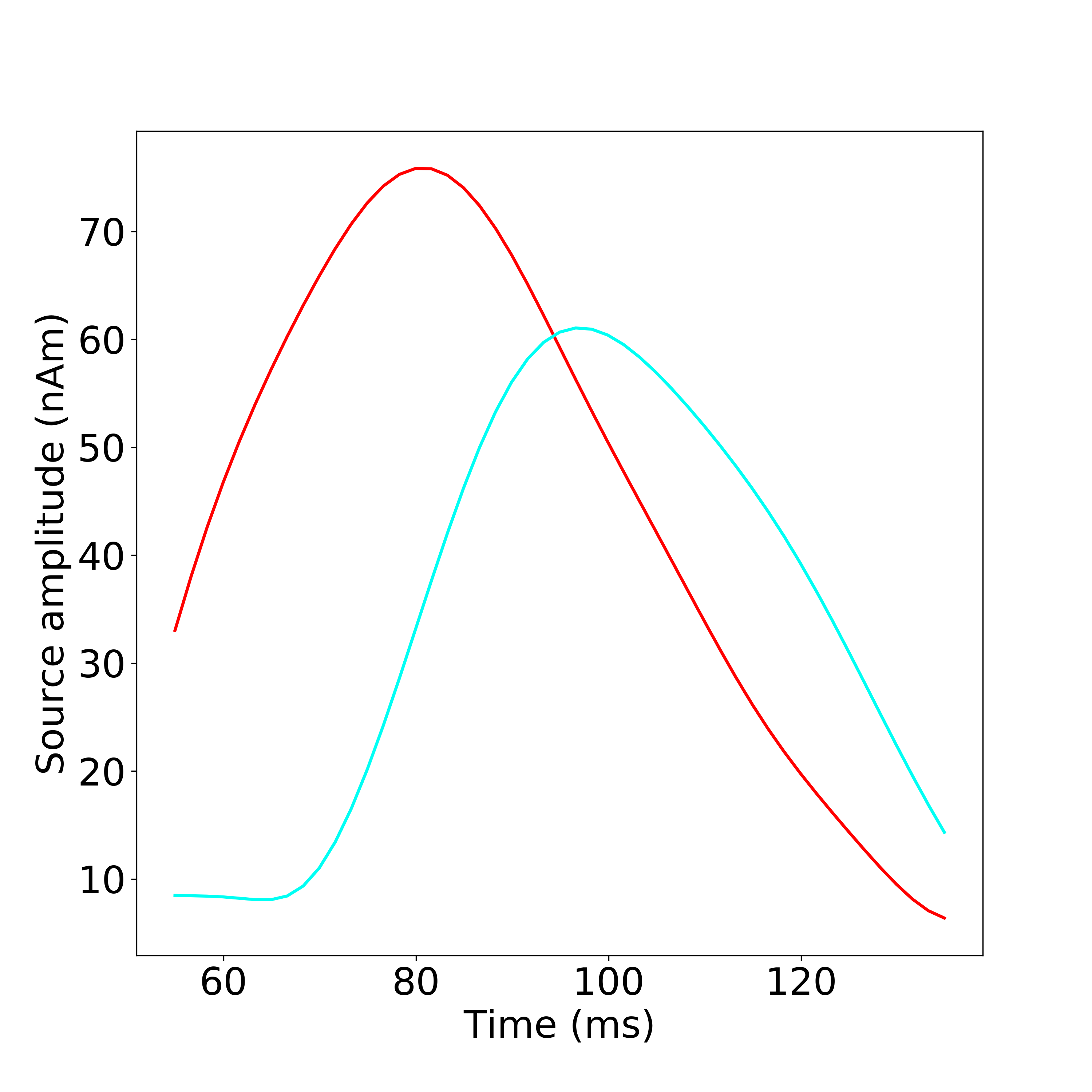}}
\label{fig:left_auditory_sesame_hyper}
\end{figure}

Figures \ref{fig:left_auditory_sesame_old} and \ref{fig:left_auditory_sesame_hyper}  show the results obtained by SESAME and h-SESAME respectively. 
In each Figure, single rows correspond to  different values of the prior scale factor, in ascending order starting from top to bottom; 
the first column shows the localization of the ECDs by means of inflated brains in which light gray indicates a gyrus, dark gray indicates a sulcus
and the auditory cortices are pictured in red;
in the second column the time course of the estimated sources is plotted, with the same color code as in the first column. 

The souce configuration estimated by SESAME in the case $k=1$ (Figure \ref{fig:left_auditory_sesame_old}, second row) is fairly in line with the 
literature \cite{kaiser2000right, gramfort2013time}: two ECDs have been estimated, one in the right auditory cortex and the other contralaterally very near to the auditory cortex;
moreover, the former activates before the latter, with a peak to peak latency difference between right and left cortices for the M$100$ activity that is quantified in $17\@$m\/s. 

Like in the case of synthetic data, the neuronal currents reconstructed by SESAME strongly depend on the given value of the hyperparameter $\sigma_q\@$: indeed, in the case $k=0.1$, eight weaker sources are estimated\@; in particular, in the left hemisphere the single ECD that is reconstructed for $k=1$ is now split in two and there are two more ECDs, one in the auditory cortex and another in the middle temporal area;
in the right hemisphere, the ECD estimated in the case $k=1$ is again split in two but also a little mislocalized and there are two more ECDs that are reconstructed almost in the marginal gyrus. The ECDs configuration that is estimated in the case $k=10$ is instead more similar to that corresponding to $k=1$, albeit sources are now deeper and stronger, and the one in the left hemisphere is farther from the auditory cortex.

Figure  \ref{fig:left_auditory_sesame_hyper} shows that the rescontructions of h-SESAME are very stable with respect to the given value of the hyperparameter
$\sigma_{min}\@$: in all three cases the same solution is provided, wich coincides with the one given by SESAME for $k=1$.

\section{Discussion}

We presented a hierarchical model for estimation of multiple current dipoles from M/EEG recordings. The new model generalizes previous work on the same topic, with multiple benefits: an improved estimate of the number of active sources; reduced localization error; great stability when the value of the input parameter varies in a wide range, to the extent that we can claim estimates are virtually independent on the input value. 
The h-SESAME algorithm represents one step forward in the automation of multi-dipole estimation from MEEG data.

Our work adds to the well established research field of hierarchical models, which are becoming increasingly popular in inverse problems, \cite{ganesan2014characterization,dunlop2017hierarchical, calvetti2020sparse,calvetti2019hierachical, calvetti2019brain}, imaging \cite{aguerrebere2017bayesian,balle2018variational}, and machine learning \cite{ansari2019hyperprior, hu2020coarse}. The main difference between our approach and that of most published research in inverse problems, is that here we adopt a fully Bayesian approach, i.e. we approximate the posterior distribution rather than computing only the maximum a posteriori estimate. 
Using a fully Bayesian approach can be a considerable advantage, as it provides the additional benefit of uncertainty quantification in two related but different ways: by quantifying the spread of the posterior distribution around its mode; and by detecting multiple modes of the posterior distribution generated by the non-uniquess of the inverse solution. We notice that proper uncertainty quantification should also account for errors in the forward model \cite{rimpilainen2017bayesian}, which are presently not considered in h-SESAME: future work will be devoted to include such uncertainty.

We remark that the introduction of the hyperprior can also be interpreted as using a non-Gaussian prior on the parameter of interest. While, for many years, the use of Gaussian priors has been the standard choice for computational reasons, it is becoming evident that their limitations are strong. Our results confirm the potential advantages of using non-Gaussian priors. Remarkably, the introduction of a hierarchical model often leads to a substantially increased computational cost, whenever an outer cycle on the value of the hyperparameter is required; in our implementation, on the contrary, the introduction of the hyperprior did not change the computational cost of the original algorithm, leading to substantially free-of-charge benefits.

Finally, we notice that, even though in this article we worked with a specific application, the described model and algorithm can be easily generalized and applied in different contexts where sparse solutions are sought for. For instance, in  \cite{sciacchitano2019sparse} an adaptive SMC algorithm is applied for a sparse imaging problem in astronomy. 
The common feature of the two problems is the high degree of sparsity of the solution, which is typically the superposition of a very small (less than 10) number of objects. It remains to be seen how well our approach would generalize to problems with a larger number of components.
Future work will be devoted to making the method more abstract and directly applicable to a wider range of sparse problems. 

\section*{Acknowledgements}
Numerical computations have been partially performed with an NVidia Quadro P6000 kindly provided to AS through an NVIDIA GPU grant program.
AS has been partially supported by Gruppo Nazionale per il Calcolo Scientifico.

\section*{References}

\bibliography{biblio}

\end{document}